\documentclass[fleqn,usenatbib]{mnras}

\usepackage[T1]{fontenc}
\usepackage{graphicx}
\usepackage{xcolor}
\usepackage{color}
\usepackage{hyperref}
\usepackage{notes2bib}
\usepackage{float}
\usepackage{array}
\usepackage{graphicx} 
\usepackage{multirow}
\usepackage{hhline,booktabs}
\usepackage{makecell}
\usepackage{amsmath,amssymb}
\usepackage{newtxtext,newtxmath}
\usepackage[normalem]{ulem}
\hypersetup{colorlinks=true,linkcolor=[rgb]{1.,0.2,0.2},citecolor=[rgb]{0.1,0.4,1.},filecolor=[rgb]{0.7,0.2,0.2},urlcolor=[rgb]{0.0,0.2,1.}}
\makeatletter
\@fleqnfalse
\@mathmargin\@centering
\makeatother

\def\INSPIRE{\mbox{{\tt INSPIRE}}}
\newcommand{\Reff}{$\mathrm{R}_{\mathrm{e}\,}$}
\newcommand{\Mstar}{$\mathrm{M}_{\star}\,$}
\newcommand{\Mfrac}{$f_{M^{\star}_{t\text{BB}=3}}$}

\newcommand{\kms}{km s$^{-1}$}
\newcommand{\ppxf}{\textsc{pPXF}}

\newcommand{\chiara}[1]{{\color{black}{#1}}}

\definecolor{darkgreen}{rgb}{0.09, 0.45, 0.27}
\definecolor{amber(sae/ece)}{rgb}{1.0, 0.49, 0.0}
\defcitealias{Ferre-Mateu+17}{F17}
\defcitealias{Tortora+18_UCMGs}{T18}
\defcitealias{Scognamiglio20}{S20}
\defcitealias{Spiniello20_Pilot}{\INSPIRE\, Pilot}
\defcitealias{Spiniello+21}{\INSPIRE\, DR1}
\defcitealias{DAgo23}{\INSPIRE\, DR2}
\defcitealias{Spiniello+23}{\INSPIRE\, DR3}
\defcitealias{Martin-Navarro+23}{\INSPIRE\, IV}

\title[INSPIRE VI. IMF and chemical composition]{ INSPIRE: INvestigating Stellar Population In RElics VI - The low-mass end slope of the stellar Initial Mass Function and chemical composition}

\author[M.~Maksymowicz-Maciata]{\noindent
Michalina Maksymowicz-Maciata$^{1}$\thanks{E-mail: michalina.maksymowicz.maciata@gmail.com}, 
Chiara Spiniello$^{1,2}$, 
Ignacio Mart\'in-Navarro$^{3,4}$, 
Anna Ferr\'e-Mateu$^{3,4}$, 
\and
Davide Bevacqua$^{5,6}$,
Michele Cappellari$^{1}$, 
Giuseppe D'Ago$^{7,2}$, 
Crescenzo Tortora$^{2}$, 
Magda Arnaboldi$^{8}$,
\and Johanna Hartke$^{9,10}$,
Paolo Saracco$^{5}$, 
Diana Scognamiglio$^{11}$\\
$^{1}$Sub-Dep. of Astrophysics, Dep. of Physics, University of Oxford, Denys Wilkinson Building, Keble Road, Oxford OX1 3RH, United Kingdom\\
$^{2}$INAF -  Osservatorio Astronomico di Capodimonte, Via Moiariello  16, 80131, Naples, Italy\\
$^{3}$Instituto de Astrof\'isica de Canarias, V\'ia L\'actea s/n, E-38205 La Laguna, Tenerife, Spain\\
$^{4}$Departamento de Astrofisica, Universidad de La Laguna, E-38200, La Laguna, Tenerife, Spain\\
$^{5}$INAF - Osservatorio Astronomico di Brera, via Brera 28, 20121 Milano, Italy\\
$^{6}$DiSAT, Universitá degli Studi dell’Insubria, via Valleggio 11, I-22100 Como, Italy\\
$^{7}$Institute of Astronomy, University of Cambridge, Madingley Road, Cambridge CB3 0HA, United Kingdom\\
$^{8}$European Southern Observatory,  Karl-Schwarzschild-Stra\ss{}e 2, 85748, Garching, Germany\\
$^{9}$Finnish Centre for Astronomy with ESO (FINCA), FI-20014 University of Turku, Finland\\
$^{10}$Tuorla Observatory, Department of Physics and Astronomy, FI-20014 University of Turku, Finland\\
$^{11}$Jet Propulsion Laboratory, California Institute of Technology, 4800,  Oak Grove Drive - Pasadena, CA 91109, USA}
\date{Accepted XXX. Received YYY; in original form ZZZ}

\pubyear{2024}

\begin{document}
\label{firstpage}
\pagerange{\pageref{firstpage}--\pageref{lastpage}}
\maketitle

\begin{abstract}
%   \emph{Context}  
The \INSPIRE\ project has built the largest sample of ultra-compact massive galaxies (UCMGs) at $0.1<z<0.4$ \chiara{and  obtained their star formation histories (SFHs).}   %obtained their star formation histories, and \chiara{computed} the degree of relicness (DoR), \chiara{to quantify} the fraction of stellar mass they formed at $z>2$. 
Due to their preserved very old stellar populations, relics are the perfect systems to constrain the earliest epochs of mass assembly in the Universe and the formation of massive early-type galaxies. 
The goal of this work is to investigate whether a correlation exists between the degree of relicness (DoR), quantifying the fraction of stellar mass formed at $z>2$, and the other stellar population parameters. 
We use the Full-Index-Fitting method %coupled with the \ppxf\ routine 
to fit the \INSPIRE\ spectra to single stellar population (SSP) models. This allows us to measure, \chiara{for the first time}, the low-mass end slope $\Gamma_{b}$ of the IMF, as well as stellar metallicity [M/H], [Mg/Fe], [Ti/Fe] and [Na/Fe] ratios, and study correlations between them and the DoR.  
Similarly to normal-sized galaxies, 
UCMGs with larger stellar masses have overall higher metallicities. We found a correlation between the low-mass end of the IMF slope and the DoR, that, however, breaks down for systems with a more extended SFH.%, for which the SSP assumption \chiara{does not hold}. 
An even stronger dependency is found between the IMF and the fraction of mass formed at high-z. 
At equal velocity dispersion and metallicity, galaxies with a higher DoR have a dwarf-richer IMF than that of low-DoR counterparts. This might indicate that the cosmic epoch and formation mechanisms influence the fragmentation of the star formation cloud and hence might be the explanation for IMF variations detected in massive ETGs.
\end{abstract}


%%%%%%%%%%%%%%%%%%%%%%%%%%%%%%%%%%%%%%%%%%%%%%%%%%

%%%%%%%%%%%%%%%%% BODY OF PAPER %%%%%%%%%%%%%%%%%%

\section{Introduction}

More than half of the total stellar mass in the Universe is contained in massive early-type galaxies (ETGs, e.g., \citealt{Renzini06}, sec.1), which play an essential role in the context of cosmic structure formation and evolution \citep{Blumenthal+84_nature}. They host the oldest populations of stars, thus retaining the memory of the earliest star formation activity. However, in local, giant massive ETGs, %according to the two-phase formation scenario \citep{Oser+10,Naab+14}, 
this \chiara{oldest} population is contaminated by accreted and/or later formed stars. 
%after a first rapid phase of intense star formation that forms a very compact central “bulk” mass of ETGs, their lifetimes are dominated by mergers, interactions and gas inflows, creating more younger population of stars, which greatly contaminate the information about the early Universe encoded in the older populations. 
Luckily, since galaxy interactions and mergers are stochastic phenomena, a small fraction of galaxies avoid the accretion and merging phase and retain their original stellar population and ultra-compact nature, becoming the \emph{relics} of the ancient Universe \citep{Trujillo+09_superdense, Taylor+10_compacts, Valentinuzzi+10_WINGS, Poggianti+13}. As relic galaxies can be observed at low redshifts with their structure intact, these ‘sealed time-capsules’ allow us to study the earliest cosmic epochs with the amount of detail only achievable in the nearby Universe. 
Studying relic galaxies can therefore help answer many questions about the early Universe and the formation and evolution of massive ETGs. This is the reason why, in the last few years, UCMGs in general, and relics more specifically, received a lot of attention from the scientific community. 
To date, only \chiara{a dozen} of relics have been confirmed and fully characterised at $z\sim0$ \citep{Trujillo14, Ferre-Mateu+17, Yildirim17, Comeron23, Salvador-Rusinol2021_nature}\footnote{\chiara{The number depends on the mass and size threshold used to define relics.}}. From Hubble Space Telescope high-resolution imaging, their morphology and density profiles match those found in high-z passive UCMGs \citep{Trujillo14,Ferre-Mateu+17}. From spatially resolved spectroscopy, it has been found that these three relics have large rotation velocities ($V\sim 200$ -- $300$ \kms) and very high central stellar velocity dispersion values ($\sigma_{\star}>300$ \kms). 
Finally, from a stellar population point of view, they have very peaked and extremely high-z star formation histories (SFH) and are populated by stars with super-solar metallicities (Z$\sim0.2$-$0.3$, with strong spatial gradients) and [Mg/Fe], old ages ($\sim 13$ Gyr), and a bottom-heavy (i.e. dwarf rich) Initial Mass Function (IMF) slope \citep{Martin-Navarro+15_IMF_relic}. 
%The IMF in these three objects has a fraction of low-mass stars being at least a factor of 2 larger than that found in the Milky Way \citep{Martin-Navarro+15_IMF_relic, Ferre-Mateu+17}  out to at least their effective radii. Conversely, stellar population analysis on normal-sized ETGs of similar stellar masses, suggests that the bottom-heavy IMF is concentrated only in the very centre \citep{vanDokkum17,Sarzi+18, Parikh+18, LaBarbera+19, Barbosa21}. 
%This suggests that the excess of dwarf stars originates from the first phase of the mass assembly, the only one that relics have experienced \citep{Barbosa21_letter, Smith2020ARAA}. 

To validate these results, a larger statistical sample is necessary, \chiara{whilst} enlarging the redshift boundaries outside the local Universe. This has been the goal of the \INSPIRE\ Project. Targeting 52 spectroscopically confirmed UCMGs at $0.1<z<0.4$, \INSPIRE\ has built the largest sample of spectroscopically confirmed UCMGs with measured kinematics and stellar population parameters. Of these, 38 have been classified as relics, as they formed more than 75\% of their stellar masses already by $z>2$. The survey has been presented in \citet{Spiniello20_Pilot} and \citet[][hereafter INSPIRE DR1]{Spiniello+21}. In \citet[][hereafter INSPIRE DR2]{DAgo23}, extensive tests were performed on the kinematics, deriving the stellar velocity dispersion within an aperture encapsulating 50\% of the light. However, since the spectra are fully seeing-dominated, and the seeing is much larger than the effective radii of the UCMGs, the velocity dispersion values have to be interpreted as lower limits\footnote{See Appendix A in \citet{Spiniello+21}.}. Finally, the stellar populations analysis of the entire sample has been presented in \citet[][hereafter INSPIRE DR3]{Spiniello+23}. 
\chiara{In \citet{Ferre-Mateu+17} it was suggested that three local relics could be ranked from the most extreme to the least extreme, in terms of their morphological and stellar population characteristics, hence following a degree of relicness (DoR). Motivated by this finding, and thanks to the larger statistical sample, \INSPIRE\ was able to quantify such DoR parameter, 
%In the same paper, motivated by the results in \citet{Ferre-Mateu+17} that found that the three relics could be ranked from the most extreme to the least extreme in terms of their morphological and stellar population characteristics, a degree of relicness (DoR) parameter was defined. This is 
as a dimensionless number, varying from 0 to 1.} The DoR is defined in terms of the fraction of stellar mass formed by $z=2$ (assumed to be the end of the first phase, \citealt{Zolotov15}), the cosmic time at which a galaxy has assembled 75\% of its mass and the final assembly time. According to this, galaxies with a high DoR are the most extreme relics as they assembled their stellar mass at the earliest epochs and very quickly, while low-DoR objects show a non-negligible fraction of later-formed populations and hence a spread in ages and metallicities. 
It was unambiguously found that at similar stellar masses, objects with a higher DoR have larger stellar metallicity and velocity dispersion values. However, the analysis in DR3 was done while keeping the elemental abundances fixed and assuming a universal IMF, which might not be the best assumption, given the results emerging in the last few years. Indeed, since the work by \citet{vanDokkum+10}, increasing evidence has emerged supporting a non-universal IMF varying across galaxies \citep{Treu+10, Conroy_vanDokkum12a, Cappellari+12, Cappellari+13_ATLAS3D_XX,Tortora+13_CG_SIM, Spiniello+14,Spiniello+12, Spiniello+15_IMF_vs_density, LaBarbera+13_SPIDERVIII_IMF,Martin-Navarro15_CALIFA} and spatially within single massive objects \citep{Martin-Navarro+15_IMF_variation,vanDokkum17,Parikh+18, Barbosa21}. The reason for these variations is however still debated. 

For the three local relics, a bottom-heavy (i.e., with a fraction of low-mass stars being at least a factor of 2 larger than that found in the Milky Way) IMF has been inferred up to a few effective radii \citep{Martin-Navarro+15_IMF_relic, Ferre-Mateu+17, Comeron23}.  
For normal-sized massive galaxies, instead, a similarly bottom-heavy IMF, with 
a fraction of stars with M $<0.5M_{\odot}$ a factor of two larger than the one measured in the Milky Way, is only required in the innermost region \citep{Martin-Navarro+15_IMF_variation, Sarzi+18, Parikh+18,LaBarbera+19, Barbosa21_letter}. Here, according to the two-phase formation scenario, the relic component is supposed to dominate the light budget \citep{Navarro-Gonzalez13, Pulsoni21, Barbosa21_letter}. All these observations can be explained assuming that the IMF might have been bottom-heavy during the early stages of galaxy formation when the Universe was much more dense and richer in hot gas, but only if stars formed through a very intense (${\rm SFR} \ge 10^{3} {\rm M}_{\odot} {\rm yr}^{-1}$) and very short ($\tau\sim 100$ Myr) burst \citep{Chabrier_2014,Smith2020ARA&A, Barbosa21_letter}. 
Hence, in UCMGs with high DoR, where the great majority of the stellar mass was formed during the first phase at high-z, we should be able to measure a steep IMF slope even from an integrated spectrum covering a large portion of the galaxy size.  
Indeed this is what we preliminary found in \citet{Martin-Navarro+23}, the fourth paper of the \INSPIRE\ series (hereafter INSPIRE IV). 
Stacking the UVB+VIS spectra of 5 relics and those of 5 non-relics with very similar velocity dispersion and metallicity values, we measured a systematic difference in the IMF slope which is dwarf-richer for relics. 
However, stacking spectra from different galaxies, broadened the final IMF probability density distribution (PDF), especially for non-relics, where the single stellar population (SSP) assumption could be less reliable, given the more heterogeneous and extended SFHs. 
Hence, a larger statistical sample is necessary to confirm this result. This is one of the main goals of this sixth paper of the \INSPIRE\ series.

%In this paper, we are able to constrain the slope of the IMF for each \INSPIRE\ galaxy, thanks to careful index selection and the FIF technique, described in Section~\ref{sec:FIF}.  
%If it is indeed true that stars formed early on in cosmic history through a fast starburst might be distributed with a bottom-heavy IMF \citep{Barbosa21}, then, relic galaxies should have steeper IMF slope than non-relics of similar stellar mass. 
%Indeed in \citet{Martin-Navarro+23}, the fourth paper of the \INSPIRE\ series (hereafter INSPIRE IV), a preliminary analysis has been performed on systems presented in \citetalias{Spiniello+21}. In particular, the UVB+VIS spectra of 5 relics and 5 non-relics were stacked, revealing a systematic difference in the IMF slope of these two populations. 
%However, relics are very rare and, to date, only a few have been spectroscopically confirmed \citep{Ferre-Mateu+17}. Great candidates for relics are ultra-compact massive galaxies (UCMGs) due to their compact structure and red colours suggesting a very old population of stars. The INSPIRE project therefore inspires to build the largest sample of UCMGs at the redshift range 0 < z < 0.5 to identify and study relic galaxies. The sample now consists of 52 objects introduced in the \cite{spiniello+21} (DATA RELEASE 1), \cite{DAgo23} (DATA RELEASE 2) and \cite{spiniello+23} (DATA RELEASE 3), which also present their stellar population modelling results. 
\chiara{In this work, we carefully inspect one by one the 52 \INSPIRE\ UCMGs, identifying contamination and bad pixels that could affect line-indices measurements. This allows us to 
derive the IMF slopes of each single object } and study how they relate to their DoR and other stellar population and kinematical parameters. 
The paper is organised as follows. In Section~\ref{sec:data}  we introduce the data and explain the methods and analysis of the stellar populations. Section~\ref{sec:results} presents the results and correlations between the obtained stellar population parameters. We then focus on the metallicity (Sec~\ref{sec:dor_metal}) and the low-mass end of the IMF slope (Sec.~\ref{sec:IMF}) and their correlations with the DoR, as well as with the stellar velocity dispersion and with each other (Sec.~\ref{sec:literature}). Finally, Section~\ref{sec:discussion} is reserved for discussion, while in Section~\ref{sec:conclusion}, we present our conclusions and summarise the findings of the paper. 

Throughout the paper, we assume a standard $\Lambda$CDM cosmology
with H\textsubscript{0}=69.6 km s\textsuperscript{-1 }Mpc\textsuperscript{-1}, $\Omega$\textsubscript{$\Lambda$}=0.714 and $\Omega$\textsubscript{M}=0.286 \citep{Bennett14}.

\section{Data and analysis}
\label{sec:data}

\subsection{Observations}
This work uses the sample of 52 UCMGs from the \INSPIRE\ project, with data collected as part of an ESO Large Programme (LP, ID: 1104.B-0370, PI: C.~Spiniello). The programme started in P104 (October 2019) and was completed in March 2023, delivering high SNR spectra ($20\le \rm{SNR} \le80$ per \AA), from the UVB to the NIR, with the X-Shooter spectrograph (XSH, \citealt{Vernet11}). We refer the reader to \citetalias{Spiniello+21} and \citetalias{Spiniello+23} for a comprehensive description of the sample selection and characteristics.

In this paper, we use the one-dimensional (1D) \INSPIRE\ spectra extracted at the R50 radius. 
This radius is obtained considering the surface brightness profiles of the 2D spectra (for each single observation block) and integrating them up to the aperture that encapsulates 50\% of the total light. However, because ground-based observations are seeing-limited, and the seeing is much larger than the galaxies' size, the aperture contains a mix of light from inside and outside the real effective radius. We refer the readers to \citetalias{Spiniello+21} for a more detailed description of the 1D extraction, which however plays an almost negligible role in the relic confirmation. 

Moreover, we only consider the rest-framed and telluric-corrected combined UVB+VIS spectra smoothed at a final resolution of FWHM$=2.51$\AA, matching that of the SSP models we use for the fitting. Our choice of not including the NIR  is two-fold. First, the telluric contamination on the spectra in this wavelength range is much larger than the one affecting the VIS. Secondly, a detailed treatment of several individual element abundances, currently not included in the MILES models, is critical to properly fit the observed strengths of infrared absorption features, especially regarding Carbon-sensitive indices, which are systematically underestimated by the current SSP models \citep{Eftekhari22}.

The characteristics of the \INSPIRE, including coordinates, redshift, DoR, velocity dispersion values, and SNR inferred from the UVB and the VIS spectra, are reported in Table~\ref{tab:tab1}. \chiara{The SNR of the spectra cover a wide range and it is generally higher in the optical, as expected for red galaxies with evolved stellar populations. }
In the last two columns of the table, we list the stellar masses and size of the 52 objects, taken from \citet{Tortora+18_UCMGs} and \citet{Scognamiglio20}\footnote{The listed effective radii are the median value between the single-band ones inferred from $g$, $r$ and $i$. These are obtained by fitting a point-spread function (PSF) convolved S\'ersic profile to the images using the code \textsc{2dphot} \citep{LaBarbera_08_2DPHOT}.}.

\begin{table*}
\centering
\caption{The \INSPIRE\ sample. From left to right we give the coordinates, redshift, DoR, stellar velocity dispersion, SNR (per \AA) in the UVB and VIS arms, the stellar mass inferred from SED fitting in the $ugri$ bands, and the median effective radius in kpc \chiara{(computed from the measurements in $g$, $r$ and $i$ bands)}. These two last quantities are taken from \citet{Tortora+18_UCMGs} and \citet{Scognamiglio20}, while all the others are taken from \citetalias{Spiniello+23}. Finally, in the last column, we report whether the galaxy is in the \textit{Golden Sample}, as defined in Sec.~\ref{sec:FIF}}. 

\label{tab:tab1}
\begin{tabular}{cccccccccc|c}
\hline \\[-1em]
\multicolumn{1}{c}{ID} &
\multicolumn{1}{c}{RAJ2000} &
\multicolumn{1}{c}{DECJ2000} &
\multicolumn{1}{c}{z} &
\multicolumn{1}{c}{DoR} &
\multicolumn{1}{c}{$\sigma_{\star}$ [km/s]} &
\multicolumn{1}{c}{SNR UVB} &
\multicolumn{1}{c}{SNR VIS} &
\multicolumn{1}{c}{$\langle\mathrm{R}_{\mathrm{e}}\rangle\,$ [kpc]} &
\multicolumn{1}{c|}{M$_{\star}$ ($10^{11}$M$_{\odot}$)} &
\multicolumn{1}{c}{\textit{Golden}} \\\\[-1em]
\hline
  J0211-3155 &  32.8962202  & -31.9279437  & 0.3012 & 0.72  & $245\pm25$ & 13.9 & 46.7   & 1.07  & 0.88 & yes\\
  J0224-3143 &  36.0902655  & -31.7244923  & 0.3839 & 0.56  & $283\pm14$ & 20.9 & 71.2   & 1.55  & 2.71 & yes\\
  J0226-3158 &  36.5109217  & -31.9810149  & 0.2355 & 0.12  & $185\pm19$ & 22.9 & 58.7   & 1.32  & 0.69 & yes\\
  J0240-3141 &  40.0080971  & -31.6950406  & 0.2789 & 0.43  & $216\pm22$ & 17.9 & 54.5   & 0.81  & 0.98 & no\\
  J0314-3215 &  48.5942558  & -32.2632678  & 0.2874 & 0.42  & $178\pm9 $ & 20.7 & 54.6   & 0.66  & 1.00 & yes\\
  J0316-2953 &  49.1896388  & -29.8835868  & 0.3596 & 0.40  & $192\pm19$ & 14.3 & 46.4   & 1.02  & 0.87 & yes\\
  J0317-2957 &  49.4141028  & -29.9561748  & 0.2611 & 0.51  & $187\pm19$ & 20.1 & 51.7   & 1.05  & 0.87 & no\\
  J0321-3213 &  50.2954390  & -32.2221290  & 0.2947 & 0.37  & $211\pm11$ & 21.9 & 66.4   & 1.37  & 1.23 & yes\\
  J0326-3303 &  51.5140585  & -33.0540443  & 0.297  & 0.25  & $173\pm17$ & 21.4 & 54.5   & 1.43  & 0.93 & yes\\
  J0838+0052 & 129.5304520  &   0.8823841  & 0.2702 & 0.54  & $189\pm9 $ & 22.6 & 65.2   & 1.28  & 0.87 & yes\\
  J0842+0059 & 130.6665506  &   0.9899186  & 0.2959 & 0.73  & $324\pm32$ & 12.4 & 41.5   & 1.01  & 0.91 & no\\
  J0844+0148 & 131.0553886  &   1.8132204  & 0.2837 & 0.45  & $224\pm22$ & 12.9 & 45.0   & 1.14  & 0.71 & yes\\
  J0847+0112 & 131.9112386  &   1.2057129  & 0.1764 & 0.83  & $244\pm12$ & 24.6 & 77.5   & 1.37  & 0.99 & yes\\
  J0857-0108 & 134.2512185  &  -1.1457077  & 0.2694 & 0.39  & $166\pm17$ & 15.7 & 43.3   & 1.40  & 1.00 & yes\\
  J0904-0018 & 136.0518949  &  -0.3054848  & 0.2989 & 0.32  & $205\pm21$ & 12.6 & 44.3   & 1.16  & 1.30 & no\\
  J0909+0147 & 137.3989150  &   1.7880025  & 0.2151 & 0.79  & $401\pm20$ & 20.7 & 75.3   & 1.05  & 1.05 & yes\\
  J0917-0123 & 139.2701850  &  -1.3887918  & 0.3602 & 0.44  & $239\pm24$ & 12.2 & 50.3   & 1.37  & 2.19 & yes\\
  J0918+0122 & 139.6446428  &   1.3794780  & 0.3731 & 0.43  & $242\pm12$ & 17.6 & 70.2   & 1.71  & 2.26 & yes\\
  J0920+0126 & 140.1291393  &   1.4431610  & 0.3117 & 0.25  & $190\pm19$ & 17.9 & 55.6   & 1.51  & 0.98 & yes\\
  J0920+0212 & 140.2320835  &   2.2126831  & 0.28   & 0.64  & $246\pm25$ & 17.0 & 55.0   & 1.48  & 1.03 & no\\
  J1026+0033 & 156.7231818  &   0.5580980  & 0.1743 & 0.29  & $225\pm11$ & 38.9 & 113.6  & 1.02  & 1.48 & yes\\
  J1040+0056 & 160.2152308  &   0.9407580  & 0.2716 & 0.77  & $240\pm24$ & 11.5 & 46.7   & 1.29  & 0.93 & yes\\
  J1114+0039 & 168.6994335  &   0.6510299  & 0.3004 & 0.40  & $181\pm18$ & 19.5 & 54.0   & 1.52  & 1.62 & no\\
  J1128-0153 & 172.0885023  &  -1.8890642  & 0.2217 & 0.34  & $192\pm10$ & 21.1 & 69.2   & 1.27  & 1.30 & yes\\
  J1142+0012$^{*}$ & 175.7023296  &   0.2043419  & 0.1077 & 0.18  & $129\pm6 $ & 57.9 & 124.1  & 1.40  & 0.84 & no\\
  J1154-0016 & 178.6922829  &  -0.2779248  & 0.3356 & 0.11  & $163\pm16$ & 16.6 & 42.8   & 1.06  & 0.64 & yes\\
  J1156-0023 & 179.2186145  &  -0.3946596  & 0.2552 & 0.30  & $177\pm18$ & 22.6 & 60.9   & 1.04  & 1.39 & no\\
  J1202+0251 & 180.5132277  &   2.8515451  & 0.3298 & 0.36  & $165\pm17$ & 14.7 & 45.9   & 1.49  & 0.68 & yes\\
  J1218+0232 & 184.7355807  &   2.5449139  & 0.308  & 0.45  & $171\pm17$ & 14.6 & 42.0   & 1.40  & 0.93 & yes\\
  J1228-0153 & 187.0640987  &  -1.8989049  & 0.2973 & 0.39  & $191\pm10$ & 23.2 & 70.1   & 1.61  & 1.15 & yes\\ 
  J1402+0117 & 210.7400749  &   1.2917747  & 0.2538 & 0.31  & $166\pm25$ & 12.4 & 34.0   & 0.68  & 0.66 & yes\\
  J1411+0233 & 212.8336012  &   2.5618381  & 0.3598 & 0.41  & $217\pm11$ & 24.1 & 73.2   & 1.07  & 1.55 & yes\\
  J1412-0020 & 213.0038281  &  -0.3440699  & 0.2783 & 0.61  & $339\pm51$ & 10.5 & 31.1   & 1.42  & 1.20 & no\\
  J1414+0004 & 213.5646898  &   0.0809744  & 0.303  & 0.36  & $205\pm31$ & 11.7 & 37.4   & 1.42  & 1.18 & yes\\
  J1417+0106 & 214.3685124  &   1.1073909  & 0.1794 & 0.33  & $203\pm10$ & 39.9 & 107.7  & 1.48  & 0.91 & yes\\
  J1420-0035 & 215.1715599  &  -0.5864629  & 0.2482 & 0.41  & $209\pm31$ & 13.4 & 39.8   & 1.35  & 0.99 & yes\\
  J1436+0007 & 219.0481314  &   0.1217459  & 0.221  & 0.33  & $193\pm19$ & 21.1 & 67.2   & 1.40  & 1.15 & yes\\
  J1438-0127 & 219.5218882  &  -1.4582727  & 0.2861 & 0.78  & $218\pm22$ & 17.9 & 59.4   & 1.20  & 0.88 & yes\\
  J1447-0149 & 221.9657402  &  -1.8242806  & 0.2074 & 0.38  & $187\pm9 $ & 24.7 & 64.7   & 1.51  & 0.86 & no\\
  J1449-0138 & 222.3504660  &  -1.6459975  & 0.2655 & 0.60  & $192\pm29$ & 10.2 & 40.3   & 1.44  & 1.03 & yes\\
  J1456+0020 & 224.2361596  &   0.3353906  & 0.2738 & 0.17  & $194\pm29$ & 11.9 & 39.6   & 0.50  & 0.71 & no\\
  J1457-0140 & 224.3397592  &  -1.6691725  & 0.3371 & 0.47  & $203\pm30$ & 12.5 & 34.1   & 1.66  & 1.51 & yes\\
  J1527-0012 & 231.7772381  &  -0.2065670  & 0.4    & 0.38  & $237\pm36$ & 7.1  & 32.7   & 1.26  & 1.74 & yes\\
  J1527-0023 & 231.7522351  &  -0.3997483  & 0.3499 & 0.37  & $188\pm28$ & 9.0  & 30.4   & 1.12  & 1.15 & yes\\
  J2202-3101 & 330.5472803  &  -31.018381  & 0.3185 & 0.48  & $221\pm22$ & 13.1 & 47.6   & 1.45  & 1.10 & yes\\
  J2204-3112 & 331.2228147  &  -31.200261  & 0.2581 & 0.78  & $227\pm23$ & 14.4 & 54.1   & 1.39  & 0.90 & yes\\
  J2257-3306 & 344.3966471  &  -33.114445  & 0.2575 & 0.27  & $185\pm19$ & 17.8 & 40.0   & 1.18  & 0.93 & no\\
  J2305-3436 & 346.3356634  &  -34.603091  & 0.2978 & 0.80  & $295\pm30$ & 14.3 & 46.8   & 1.29  & 0.86 & no\\
  J2312-3438 & 348.2389042  &  -34.648591  & 0.3665 & 0.36  & $221\pm11$ & 32.1 & 72.4   & 1.25  & 1.34 & yes\\
  J2327-3312 & 351.9910156  &  -33.200760  & 0.4065 & 0.06  & $227\pm11$ & 19.6 & 72.8   & 1.51  & 1.57 & yes\\
  J2356-3332 & 359.1261248  &  -33.533475  & 0.3389 & 0.44  & $162\pm24$ & 11.5 & 34.2   & 1.06  & 0.98 & yes\\
  J2359-3320 & 359.9851685  &  -33.333583  & 0.2888 & 0.71  & $267\pm27$ & 15.6 & 49.1   & 1.04  & 1.07 & yes\\ \\[-1em]
\hline\end{tabular}
%\begin{flushright} $^*$These objects have been excluded from the analysis because they have $\ge25$\% of stars with ages younger than 1 Gyr and this breaks the SSP-approximation. \misia{should we remove this footnote? J2327-3312 would then be golden and J1142 (the figures do already include them)}
%\end{flushright}
\end{table*}

\subsection{Stellar population models}
We use the MILES SSP models developed and described in \citet{Vazdekis15} with BaSTI theoretical isochrones %\footnote{\href{http://www.oa-teramo.inaf.it/BASTI}{http://www.oa-teramo.inaf.it/BASTI}.} 
\citep{Pietrinferni04, Pietrinferni06}. 
The MILES SSPs are based on the MILES empirical stellar library\footnote{Publicly available at http://miles.iac.es.} \citep{Sanchez-Blazquez+06, Falcon11}, covering the wavelength range [3525-7500] \AA.  
The SSPs cover a broad range of stellar ages, ranging from 30 Myr to 14 Gyr, and sampled at logarithmic steps, total stellar metallicities in the range $-2.27<$[M/H]$<+0.40$, two values of the [$\alpha$/Fe] ratios (solar and super-solar, $+$0.4 dex) and a suite of stellar IMF slopes. 
For this paper, we use the Bimodal IMF parametrisation defined in \citet{Vazdekis96}. This is described by two power-law regimes with a break at 0.6 M$_{\odot}$, where the low-mass slope is fixed at 1.3 \citep{Salpeter55}, while the high-mass slope ($\Gamma_{b}$) is free to vary between 0.3 and 3.5. 

Since degeneracies exist between variation in the IMF slope, especially at its low-mass end, and single elemental abundances \citep[e.g.,][]{Spiniello+14}, we need to take this into account. 
In particular, Sodium (Na) and Titanium (Ti) absorption features have been extensively used to measure the IMF in massive ETGs\citep{Conroy_vanDokkum12b, Ferreras+13, Spiniello+12,Spiniello+14,Spiniello+15_NGC4697, Parikh+18, Sarzi+18,LaBarbera+19}. 
To model the variation of Na and Ti, we use response functions of these two elements, computed using the SSP models of \citet{Conroy_vanDokkum12a}, hereafter CvD. 
We follow similar prescriptions to these given in \citet{Spiniello15}. In particular, the response functions are computed by simply taking the ratio between two CvD spectra, with the same age, same IMF, and metallicity, one with solar abundances and one with different [Na/Fe] or [Ti/Fe]. In this way, we are able to isolate the effect of changing the elemental abundance from the effect of changing other stellar population parameters. Specifically, in this case, we compute response functions for a 13.5 Gyr population, which is sufficiently similar to the expectations for our sample given the weak age dependence of elemental abundance corrections \citep[e.g.][]{Vazdekis15}. The conclusions presented here will not change if a slightly younger age (e.g. 10-11 Gyr) would have been used to obtain the response functions and this is enough to draw conclusions on the elemental abundances in relics.  

\begin{figure*}
 \centering \includegraphics[width=1\textwidth]{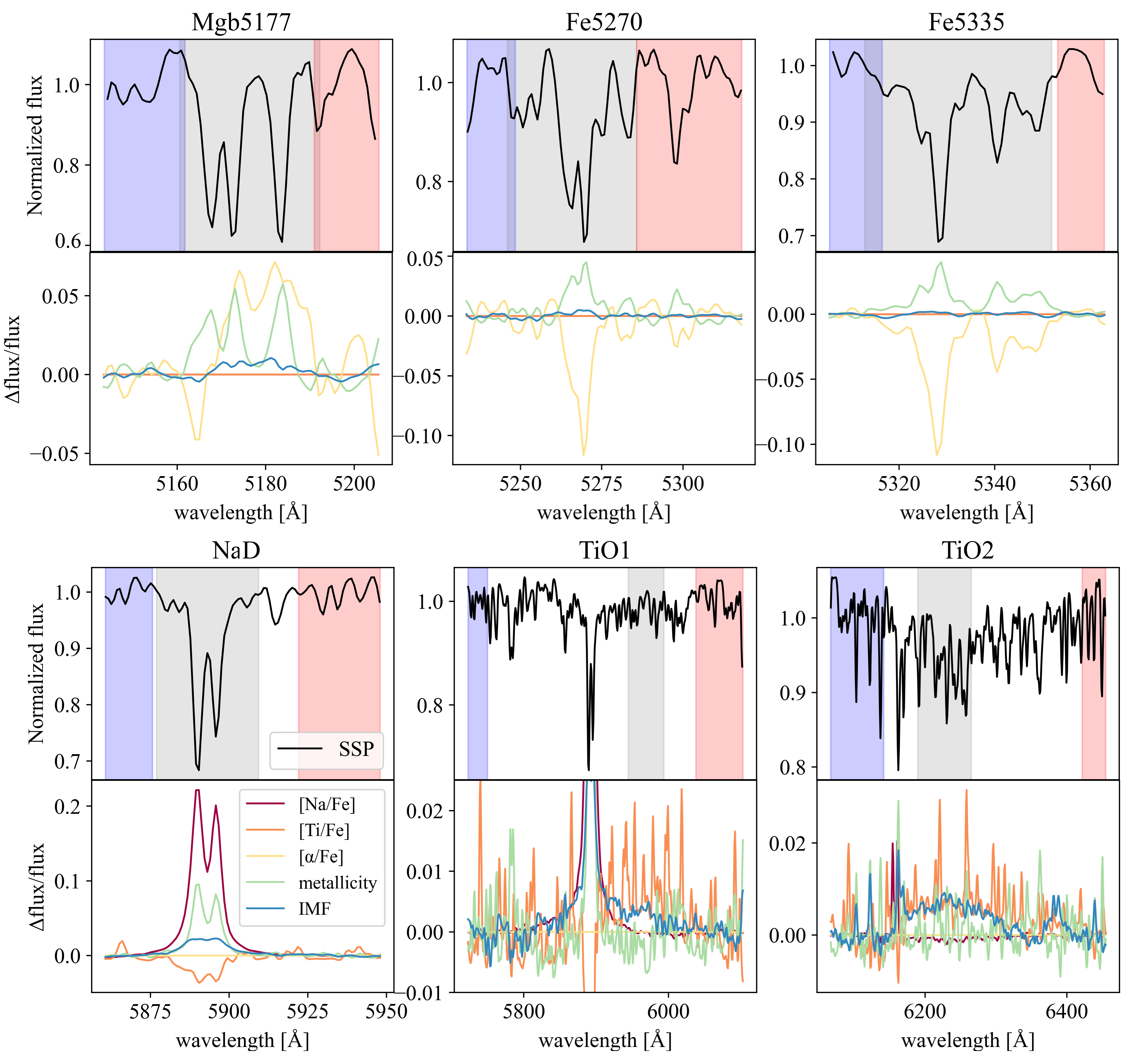}
\caption[width=0.5\textwidth]{The six indices constituting our fiducial set: Mgb, Fe5270 and Fe5335 (top line) and NaD, TiO1 and TiO2 (bottom line). For each index, the top panel shows the index (grey shaded region) and blue and red bands (blue shaded regions) around which the flux is normalised. 
%The top panels show the Mg2 (left), Fe5270 (middle) and the NaD (right) spectral features, normalized using the index pseudo-continua (blue shaded regions). In the FIF approach, every pixel within the central bandpass (grey area) is fitted to obtain the stellar population parameters. 
The black lines correspond to model spectra of solar metallicity [M/H] = 0, [Mg/Fe] = 0, [Ti/Fe] = -0.3, [Na/Fe] = -0.3 and IMF slope of $\Gamma$\textsubscript{B} = 1.3. The bottom panels show the relative change in the spectrum after varying different stellar population parameters: the IMF slope was varied by $\Delta\Gamma$\textsubscript{B} = 1, the metallicity by 0.4 dex and \chiara{the different abundance ratios by 0.6 dex each.  }}
\label{fig:indices}

\end{figure*}

\subsection{Full-Index Fitting}
\label{sec:FIF}
The core of the modelling in this work revolves around the Full-Index Fitting method (FIF, \citealt{Martin-Navarro+19,Martin-Navarro+21}), which is a hybrid approach between a more standard line-strength analysis and full-spectrum fitting. Instead of calculating the equivalent widths of key absorption features or fitting every pixel across a wide wavelength range, this method fits every pixel within the selected key absorption features to the models, after normalising the continuum using the index definition. 
This hybrid approach has lower SNR requirements than pure line-strength, reduced computational time than full-spectrum fitting \citep{Martin-Navarro+19}, and allows to focus on well-tested and studied spectral regions, where the information about specific stellar population properties is concentrated.  
%In this way, the hybrid approach gets the best of both methods: lower SNR requirements, reduced computational time, and a focus on well-tested and studied spectral regions, where the information about the stellar population properties is concentrated. 

We use FIF in combination with the Penalised Pixel-fitting software\footnote{\url{https://pypi.org/project/ppxf/}} (\ppxf; \citealt{Cappellari04,Cappellari17, Cappellari23}), already used in previous \INSPIRE\ publications. Specifically, we first run \ppxf\ on the UVB+VIS combined and smoothed spectra, computing the stellar velocity dispersion, as well as the light-weighted stellar age (in logarithmic scale) and metallicity, and a first guess of the IMF slope. %\sout{We note that the publicly available version of the code only works in the age-metallicity space.} 
\chiara{We note that we have customised the publicly available version of the code to let it constrain also the slope of the IMF}, but we keep elemental abundance ratios fixed to solar and we do not fit for them.   \chiara{Moreover, we stress to the readers in \citetalias{Spiniello+23} we performed two separate runs, one for kinematics and one for stellar population constraints. Here instead, we perform a single run. We, therefore, believe that, although we obtain the stellar velocity dispersion as a bi-product, we believe that the values computed in DR3 are more trustable, as the \ppxf\ code was optimised for kinematics (e.g., using an additive polynomial and performing tests on the systematics and random errors). Hence we list the DR3 values in Table~\ref{tab:tab1} and plot and use these values throughout the paper. We nevertheless show in Appendix~\ref{app:comparisonDR3} that the agreement between the two measurements is good (see the rightmost panel of Fig.~\ref{fig:dr3comp}).  }

\chiara{Following the same line of thoughts, we did not re-derive the DoR here based on the new ages and metallicities computed with \ppxf. Indeed, we did not perform any regularisation during the fit (see \citetalias{Spiniello+23} for details), which often revealed signs of younger ages, when there, as in the case of systems with low  DoR. We, therefore, believe that the DoR computed in \citetalias{Spiniello+23} is more correct and use that for the remainder of the paper. }

%At this stage, we use SSP models with varying IMF and obtain a first guess for its slope, but we keep elemental abundance ratios fixed to solar. 

At this point, we input the \chiara{best-fitting} age and the velocity dispersion values to the FIF routine as fixed values and obtain an estimate for the stellar metallicity, Mg, Na and Ti abundances, and the IMF slope ($\Gamma_b$) with their associated uncertainties. 
\chiara{We use [$\alpha$/Fe] as a proxy of [Mg/Fe]. This is motivated by the fact that even though the models are built at varying [$\alpha$/Fe] the $\alpha$ element we are most sensitive to in the wavelength range we use, is Magnesium. Indeed, although TiO bands also depend on [$\alpha$/Fe], this effect is counter-balanced by the sensitivity to Carbon and by the fact that we also allow for [Ti/Fe] variation. Therefore, hereafter, we will always quote and refer to [Mg/Fe] ratios.}

Note that, although the model grid is in principle discrete, the best-fitting stellar population values and uncertainties are estimated by linearly interpolating the closest nodes in the grid. 

Following \citet{Martin-Navarro+19}, the age is kept fixed for this second step to the one obtained with pPXF, in order to circumvent the effect of [C/Fe] on the strength of the H$_\beta$ feature \citep{Conroy_vanDokkum12a}. In practice, this approach minimises any potential [C/Fe]-age degeneracy \citep{Martin-Navarro+19}. We have nevertheless performed a FIF run letting the age free to vary and including one or more age-sensitive lines (e.g., Balmer lines) and another run where we fix the age to that obtained in DR3. More details on these tests are given in Appendix~\ref{app:testing}. %to constrain the stellar age too, but we note that in this case, we cannot control for the [C/Fe]-age degeneracy \citep{Martin-Navarro+19}. 

In principle, the larger the number of indices fitted, the better the constraining power should be \citep{Spiniello+14}. 
However, some of the optical spectral indices for several galaxies in the \INSPIRE\ sample are contaminated by residuals of skylines and bad pixels, rendering them unfit to use. 
Hence, we need to select the minimum number of indices that allow us to break the degeneracy between the stellar population parameters while maximising the number of galaxies without any signs of contamination. 
After extensive testing (see Appendix~\ref{app:testing}), we selected 6 optical indices, namely Mgb, Fe5270, Fe5335, NaD, TiO1, TiO2\footnote{The indices definition, including also blue and red band-passes are given in Table~\ref{tab:indices}, in  Appendix~\ref{app:indices}}, defined around stellar absorption features mainly coming from 4 different chemical species: Magnesium, Iron, Titanium, and Sodium. 
These are shown in the top panels of Figure~\ref{fig:indices}. In the corresponding bottom panels, we showcase how they respond to changes to the studied stellar population parameters. 
The combination of these six indices 
allows us to break the degeneracies among the stellar population parameters and enables us to correctly measure stellar metallicity, IMF slope, Mg, Na and Ti abundances. 

Starting from these six indices, and checking galaxy by galaxy whether the spectra present bad pixels and contaminated regions affecting the calculation of the equivalent width, we define a \textit{Golden Sample} of 39 galaxies, for which none of the six selected features is contaminated \chiara{(see last column of Table~\ref{tab:tab1})}.%, and the stellar populations are old enough that the assumption of an SSP is valid. 

%\anna{I  START  commenting again  from this part, for previous comments and corrections  check the older  version...I corrected typos too, so you  might want to  copy  paste, I also fixed  the  \AA not showing}
In Appendix~\ref{app:testing}, we present a series of tests aimed at checking whether the results of the FIF depend %greatly depend 
on the spectral indices chosen to be fitted to the models, and which parameters are let free in the run. %In particular, we also test the inclusion of one or more age-sensitive lines (e.g., Balmer lines) to constrain the stellar age too, but we note that in this case, we cannot control for the [C/Fe]-age degeneracy \citep{Martin-Navarro+19}. 
The main conclusion of these tests is that the only two significant differences in the inferred stellar population parameters arise when i) changing the Mg index used for the fit (from Mgb to Mg2), or ii) letting the age free to vary, adding a Balmer line to constrain it. 
However, importantly, the inference on the IMF is run-independent and therefore very robust and the results are almost completely unchanged for extreme relics, while a larger variation is found for galaxies with lower DoR. 

\begin{figure}
    \centering
    \includegraphics[width=\linewidth]{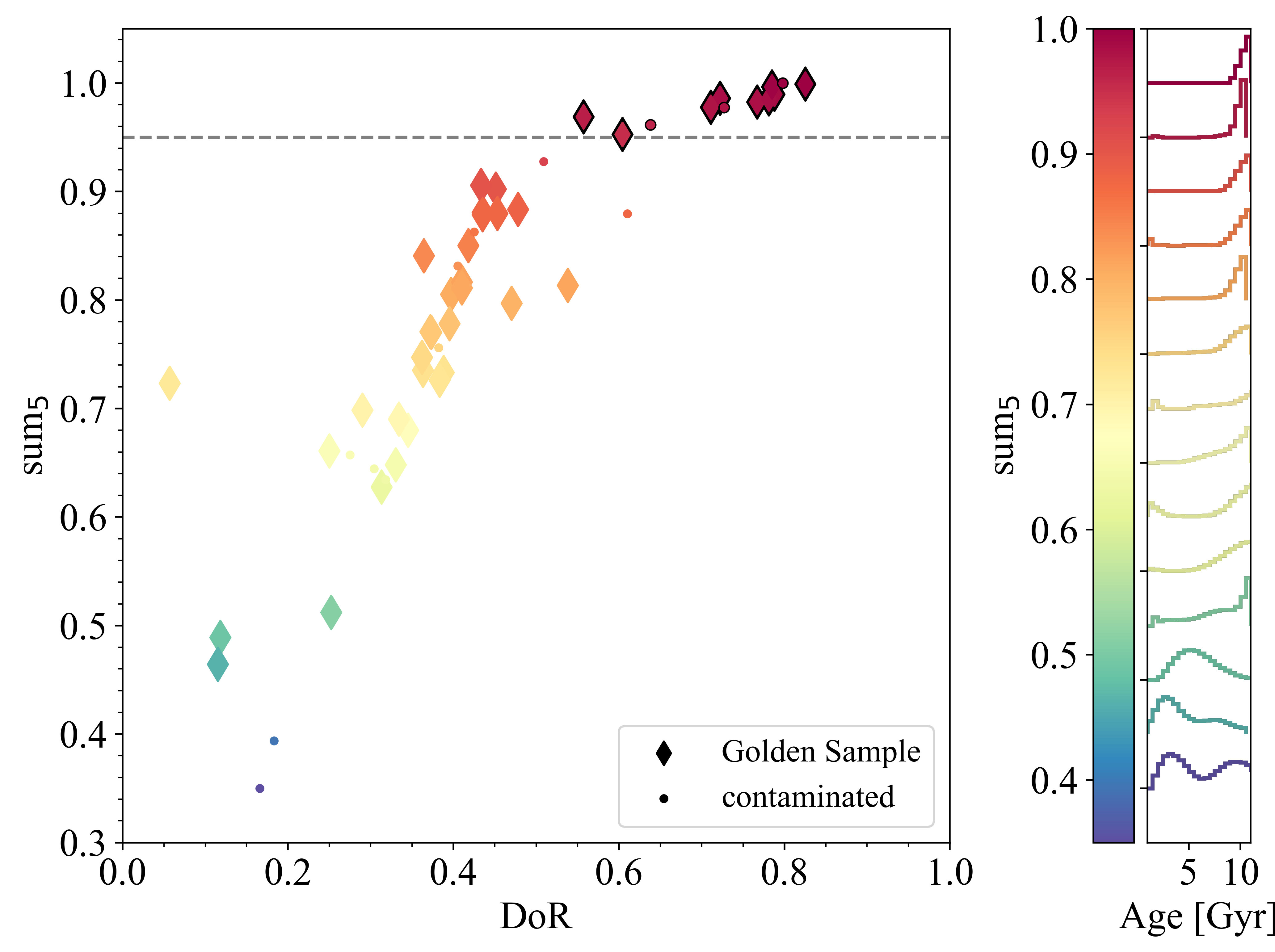}
    \caption{The degree of relicness against the maximum sum of the weights held by 5 neighbouring bins in the age distribution histograms. Dashed line marks the 0.95 threshold of the sum of weights. On the right-most panel a sample of age-distributions is presented for a quick reference. }
\label{fig:nbin_vs_dor}
\end{figure}

\subsection{Testing the SSP assumption}
\label{sec:sspness}
We remind the readers that \ppxf\ uses a series of SSPs to derive the star formation history of a galaxy. For extreme relics, when the entire totality of the stellar mass has been formed quickly at very early cosmic time, it is expected that the code only uses a low number of very old SSPs. At decreasing DoR, the fraction of mass formed during the first phase of the formation scenario becomes smaller and the star formation history more extended. Hence the number of SSP models, with different ages and metallicities, used for the fit increases. This also means that the assumption that the star formation history can be reconstructed by combining SSP is less solid. 
In this section, we investigate the spread in age that \ppxf\ attributes to each galaxy and how this correlates with the DoR \chiara{obtained in \citetalias{Spiniello+23}}. We believe that this number represents a qualitative way to measure the validity of the SSP assumption: the larger the number of age bins is, the worse an SSP approximation is for a given galaxy. 

Figure~\ref{fig:nbin_vs_dor} presents how the DoR correlates with the maximum sum of the weights of 5 neighbouring bins of the age-distribution histogram \chiara{(sum$_{5}$). To compute this number, we start by taking all the non-zero weights that \ppxf\ has attributed to the array of SSP models (with different ages and metallicities). Then, marginalising over metallicity, we produce an age-distribution histogram. From it, we calculate the sum of weights attributed by \ppxf\ to every five neighbouring age-bins (corresponding to 2.5 Gyr, given that the models span an age grid of 0.5 Gyr), and check which group of five holds the greatest sum of their weights.} 
As expected a clear correlation is found. Moreover, we also note that a plateau is reached for the sum of weights $>0.95$. 
In the right panel of the same figure, we sketch the age distribution for different age bins. The threshold of 0.95 corresponds to a very peaked distribution of ages above 9 Gyr, without any second peak at lower redshift (younger ages). \chiara{Below this number, the age histograms start to be broader and peaks at younger ages appear and therefore the SSP-assumption becomes less valid. } %We therefore conclude that the assumption of SSP is certainly valid for objects that have the maximum sum of weights of 5 neighbouring bins greater than 0.95.
\chiara{We note that even though the sum$_{5}$ strongly correlated with the DoR for similar, intermediate DoR (0.4-0.6), different galaxies can have different sum$_{5}$.} In the remainder of this paper we will use this threshold to derive relations between stellar population parameters that might hold only for stars formed through a quick star formation episode at early cosmic time. 
We note that the threshold choice is arbitrary and different relations could be found based on different definitions, \chiara{however we believe it to be informative to check whether} the correlations hold especially for the most strict age-distribution constraints. 
Furthermore, we acknowledge that the sum of the weights expresses a concept which is very similar to that measured with the DoR, but we note that the latter depends on the fitting assumptions within \ppxf, which are different between this paper and the \citetalias{Spiniello+23} (see Appendix~\ref{app:comparisonDR3} for more details).

\section{Stellar population analysis results}
\label{sec:results}
This Section presents the main results of the FIF analysis in terms of stellar population parameters and their relation with the DoR, obtained in \citetalias{Spiniello+23}. A direct comparison with the results obtained in DR3 is presented in Appendix~\ref{app:comparisonDR3}. %and compares UCMGs with normal-sized galaxies of similar stellar masses. 
%The IMF slope will instead be covered in Section~\ref{sec:IMF}.
When not otherwise specified, we will focus on results obtained from the fiducial run, which are listed in Table \ref{tab:stel_pop_results}.

\begin{table}
\centering
\caption{Stellar population results from FIF for the fiducial model.}
\begin{tabular}{@{}cccccc@{}}
\toprule
ID         & $\Gamma_b$                    & [M/H]                   & [Mg/Fe]          & [Ti/Fe]                 & [Na/Fe]                 \\ \midrule
J0211-3155 & 1.71$^{+0.30}_{-0.32}$ & -0.06$^{+0.06}_{-0.06}$   & 0.07$^{+0.06}_{-0.04}$            & -0.23$^{+0.09}_{-0.06}$     & 0.06$^{+0.07}_{-0.08}$      
\\\\[-1em]
J0224-3143 & 0.88$^{+0.32}_{-0.35}$ & 0.14$^{+0.04}_{-0.03}$    & 0.22$^{+0.05}_{-0.04}$            & -0.26$^{+0.04}_{-0.02}$     & 0.28$^{+0.01}_{-0.03}$      
\\\\[-1em]
J0226-3158 & 1.54$^{+0.39}_{-0.44}$ & -0.47$^{+0.05}_{-0.05}$   & 0.17$^{+0.07}_{-0.07}$            & -0.24$^{+0.15}_{-0.09}$     & 0.18$^{+0.05}_{-0.05}$      
\\\\[-1em]
J0240-3141 & 1.75$^{+0.44}_{-0.41}$ & 0.36$^{+0.03}_{-0.04}$    & 0.08$^{+0.04}_{-0.04}$            & -0.28$^{+0.04}_{-0.02}$     & 0.05$^{+0.08}_{-0.10}$      
\\\\[-1em]
J0314-3215 & 1.14$^{+0.41}_{-0.39}$ & 0.01$^{+0.07}_{-0.05}$    & 0.01$^{+0.02}_{-0.01}$            & -0.20$^{+0.13}_{-0.09}$     & 0.24$^{+0.04}_{-0.05}$      
\\\\[-1em]
J0316-2953 & 1.76$^{+0.38}_{-0.44}$ & -0.33$^{+0.10}_{-0.07}$   & 0.07$^{+0.06}_{-0.05}$            & -0.24$^{+0.08}_{-0.04}$     & 0.27$^{+0.02}_{-0.04}$      
\\\\[-1em]
J0317-2957 & 2.55$^{+0.35}_{-0.71}$ & 0.06$^{+0.04}_{-0.03}$    & 0.23$^{+0.07}_{-0.06}$            & 0.27$^{+0.03}_{-0.05}$      & 0.12$^{+0.09}_{-0.08}$      
\\\\[-1em]
J0321-3213 & 1.24$^{+0.30}_{-0.30}$ & 0.11$^{+0.06}_{-0.05}$    & 0.07$^{+0.04}_{-0.04}$            & -0.26$^{+0.06}_{-0.02}$     & 0.00$^{+0.07}_{-0.07}$      
\\\\[-1em]
J0326-3303 & 1.80$^{+0.29}_{-0.36}$ & -0.11$^{+0.05}_{-0.05}$ & 0.12$^{+0.06}_{-0.06}$ & -0.19$^{+0.14}_{-0.09}$ & -0.04$^{+0.06}_{-0.06}$ 
\\\\[-1em]
J0838+0052 & 2.20$^{+0.29}_{-0.40}$ & 0.07$^{+0.03}_{-0.02}$    & 0.15$^{+0.04}_{-0.04}$            & 0.28$^{+0.02}_{-0.04}$      & -0.03$^{+0.05}_{-0.06}$     
\\\\[-1em]
J0842+0059 & 2.80$^{+0.26}_{-0.32}$ & 0.36$^{+0.03}_{-0.04}$    & 0.26$^{+0.06}_{-0.06}$            & -0.18$^{+0.16}_{-0.10}$     & 0.27$^{+0.02}_{-0.05}$      
\\\\[-1em]
J0844+0148 & 2.70$^{+0.25}_{-0.26}$ & -0.19$^{+0.05}_{-0.05}$   & 0.16$^{+0.06}_{-0.06}$            & 0.01$^{+0.16}_{-0.18}$      & 0.24$^{+0.04}_{-0.06}$      
\\\\[-1em]
J0847+0112 & 2.44$^{+0.25}_{-0.37}$ & 0.09$^{+0.06}_{-0.06}$    & 0.20$^{+0.05}_{-0.05}$            & 0.12$^{+0.20}_{-0.16}$      & 0.26$^{+0.03}_{-0.05}$      
\\\\[-1em]
J0857-0108 & 0.79$^{+0.41}_{-0.32}$ & -0.02$^{+0.05}_{-0.05}$   & 0.36$^{+0.03}_{-0.05}$            & 0.18$^{+0.10}_{-0.16}$      & -0.20$^{+0.06}_{-0.05}$     
\\\\[-1em]
J0904-0018 & 1.79$^{+0.37}_{-0.42}$ & 0.00$^{+0.06}_{-0.06}$    & 0.13$^{+0.07}_{-0.06}$            & -0.20$^{+0.13}_{-0.07}$     & 0.04$^{+0.07}_{-0.08}$      
\\\\[-1em]
J0909+0147 & 2.35$^{+0.23}_{-0.28}$ & 0.20$^{+0.05}_{-0.05}$    & 0.12$^{+0.05}_{-0.05}$            & -0.04$^{+0.17}_{-0.11}$     & 0.27$^{+0.02}_{-0.04}$      
\\\\[-1em]
J0917-0123 & 2.52$^{+0.44}_{-0.68}$ & 0.21$^{+0.06}_{-0.07}$    & 0.23$^{+0.06}_{-0.06}$            & -0.24$^{+0.16}_{-0.07}$     & 0.29$^{+0.01}_{-0.01}$      
\\\\[-1em]
J0918+0122 & 1.61$^{+0.33}_{-0.36}$ & -0.06$^{+0.05}_{-0.04}$   & 0.26$^{+0.05}_{-0.05}$            & -0.24$^{+0.08}_{-0.04}$     & 0.21$^{+0.05}_{-0.04}$      
\\\\[-1em]
J0920+0126 & 0.67$^{+0.27}_{-0.25}$ & -0.15$^{+0.04}_{-0.04}$   & 0.12$^{+0.05}_{-0.05}$            & -0.25$^{+0.07}_{-0.03}$     & 0.10$^{+0.04}_{-0.05}$      
\\\\[-1em]
J0920+0212 & 2.30$^{+0.37}_{-0.44}$ & 0.25$^{+0.06}_{-0.06}$    & 0.14$^{+0.05}_{-0.05}$            & -0.22$^{+0.12}_{-0.06}$     & 0.02$^{+0.10}_{-0.11}$      
\\\\[-1em]
J1026+0033 & 2.63$^{+0.13}_{-0.15}$ & 0.20$^{+0.02}_{-0.02}$    & 0.21$^{+0.03}_{-0.03}$            & -0.29$^{+0.01}_{-0.01}$     & 0.30$^{+0.00}_{-0.00}$      
\\\\[-1em]
J1040+0056 & 3.02$^{+0.18}_{-0.23}$ & 0.12$^{+0.05}_{-0.05}$    & 0.25$^{+0.05}_{-0.05}$            & 0.22$^{+0.08}_{-0.13}$      & -0.16$^{+0.09}_{-0.08}$     
\\\\[-1em]
J1114+0039 & 1.32$^{+0.46}_{-0.42}$ & -0.45$^{+0.05}_{-0.05}$   & 0.39$^{+0.00}_{-0.01}$            & -0.21$^{+0.14}_{-0.09}$     & 0.26$^{+0.03}_{-0.04}$      
\\\\[-1em]
J1128-0153 & 0.73$^{+0.26}_{-0.25}$ & 0.08$^{+0.03}_{-0.04}$    & 0.04$^{+0.03}_{-0.03}$            & -0.25$^{+0.06}_{-0.03}$     & 0.29$^{+0.01}_{-0.02}$      
\\\\[-1em]
J1142+0012 & 3.28$^{+0.01}_{-0.04}$ & -0.46$^{+0.03}_{-0.03}$   & 0.03$^{+0.04}_{-0.02}$            & -0.30$^{+0.01}_{-0.00}$     & 0.30$^{+0.00}_{-0.00}$      
\\\\[-1em]
J1154-0016 & 1.39$^{+0.50}_{-0.49}$ & 0.25$^{+0.07}_{-0.07}$    & 0.05$^{+0.05}_{-0.03}$            & 0.00$^{+0.06}_{-0.11}$      & 0.14$^{+0.06}_{-0.07}$      
\\\\[-1em]
J1156-0023 & 1.45$^{+0.29}_{-0.26}$ & 0.11$^{+0.04}_{-0.03}$    & 0.01$^{+0.02}_{-0.01}$            & -0.29$^{+0.03}_{-0.01}$     & -0.03$^{+0.05}_{-0.05}$     
\\\\[-1em]
J1202+0251 & 0.94$^{+0.35}_{-0.32}$ & -0.07$^{+0.05}_{-0.05}$   & 0.05$^{+0.05}_{-0.04}$            & -0.23$^{+0.10}_{-0.05}$     & 0.03$^{+0.06}_{-0.05}$      
\\\\[-1em]
J1218+0232 & 1.57$^{+0.47}_{-0.48}$ & -0.13$^{+0.05}_{-0.05}$   & 0.12$^{+0.06}_{-0.06}$            & 0.03$^{+0.17}_{-0.18}$      & 0.14$^{+0.07}_{-0.08}$      
\\\\[-1em]
J1228-0153 & 0.79$^{+0.37}_{-0.32}$ & 0.03$^{+0.04}_{-0.04}$    & 0.17$^{+0.04}_{-0.04}$            & -0.19$^{+0.12}_{-0.09}$     & 0.06$^{+0.05}_{-0.05}$      
\\\\[-1em]
J1402+0117 & 0.83$^{+0.44}_{-0.34}$ & -0.47$^{+0.07}_{-0.07}$   & 0.11$^{+0.10}_{-0.08}$            & -0.20$^{+0.17}_{-0.10}$     & 0.20$^{+0.06}_{-0.08}$      
\\\\[-1em]
J1411+0233 & 1.97$^{+0.33}_{-0.28}$ & -0.19$^{+0.04}_{-0.11}$   & 0.13$^{+0.05}_{-0.05}$            & -0.21$^{+0.10}_{-0.05}$     & 0.28$^{+0.01}_{-0.02}$      
\\\\[-1em]
J1412-0020 & 1.54$^{+0.36}_{-0.39}$ & -0.08$^{+0.10}_{-0.07}$   & 0.02$^{+0.03}_{-0.01}$            & -0.19$^{+0.08}_{-0.05}$     & 0.28$^{+0.01}_{-0.04}$      
\\\\[-1em]
J1414+0004 & 0.99$^{+0.44}_{-0.38}$ & -0.12$^{+0.06}_{-0.06}$   & 0.27$^{+0.06}_{-0.06}$            & -0.24$^{+0.06}_{-0.03}$     & 0.28$^{+0.01}_{-0.02}$      
\\\\[-1em]
J1417+0106 & 0.62$^{+0.27}_{-0.22}$ & -0.12$^{+0.02}_{-0.02}$   & 0.11$^{+0.03}_{-0.03}$            & 0.16$^{+0.07}_{-0.10}$      & 0.08$^{+0.03}_{-0.03}$      
\\\\[-1em]
J1420-0035 & 1.30$^{+0.43}_{-0.46}$ & -0.32$^{+0.07}_{-0.06}$   & 0.21$^{+0.09}_{-0.08}$            & -0.18$^{+0.17}_{-0.10}$     & 0.25$^{+0.03}_{-0.05}$      
\\\\[-1em]
J1436+0007 & 2.65$^{+0.28}_{-0.29}$ & -0.12$^{+0.04}_{-0.04}$   & 0.05$^{+0.04}_{-0.04}$            & 0.08$^{+0.13}_{-0.22}$      & -0.04$^{+0.08}_{-0.09}$     
\\\\[-1em]
J1438-0127 & 2.47$^{+0.24}_{-0.31}$ & 0.17$^{+0.04}_{-0.05}$    & 0.18$^{+0.04}_{-0.04}$            & 0.19$^{+0.07}_{-0.15}$      & -0.16$^{+0.08}_{-0.08}$     
\\\\[-1em]
J1447-0149 & 2.25$^{+0.50}_{-0.77}$ & 0.32$^{+0.04}_{-0.04}$    & 0.01$^{+0.01}_{-0.01}$            & 0.25$^{+0.04}_{-0.07}$      & -0.13$^{+0.09}_{-0.09}$     
\\\\[-1em]
J1449-0138 & 2.15$^{+0.40}_{-0.49}$ & -0.20$^{+0.05}_{-0.05}$   & 0.22$^{+0.06}_{-0.06}$            & 0.21$^{+0.06}_{-0.12}$      & 0.07$^{+0.07}_{-0.08}$      
\\\\[-1em]
J1456+0020 & 1.53$^{+0.85}_{-0.84}$ & -0.07$^{+0.06}_{-0.05}$   & 0.01$^{+0.02}_{-0.01}$            & 0.27$^{+0.03}_{-0.05}$      & 0.19$^{+0.06}_{-0.07}$      
\\\\[-1em]
J1457-0140 & 0.80$^{+0.32}_{-0.30}$ & -0.13$^{+0.07}_{-0.07}$   & 0.04$^{+0.05}_{-0.03}$            & -0.16$^{+0.20}_{-0.14}$     & 0.15$^{+0.08}_{-0.08}$      
\\\\[-1em]
J1527-0012 & 1.05$^{+0.48}_{-0.40}$ & -0.10$^{+0.07}_{-0.07}$   & 0.13$^{+0.10}_{-0.08}$            & 0.11$^{+0.19}_{-0.21}$      & 0.26$^{+0.03}_{-0.05}$      
\\\\[-1em]
J1527-0023 & 3.21$^{+0.06}_{-0.12}$ & -0.19$^{+0.08}_{-0.07}$   & 0.10$^{+0.09}_{-0.07}$            & -0.24$^{+0.11}_{-0.06}$     & 0.28$^{+0.01}_{-0.03}$      
\\\\[-1em]
J2202-3101 & 0.67$^{+0.39}_{-0.25}$ & 0.26$^{+0.05}_{-0.05}$    & 0.13$^{+0.04}_{-0.04}$            & 0.03$^{+0.16}_{-0.17}$      & -0.05$^{+0.08}_{-0.09}$     
\\\\[-1em]
J2204-3112 & 3.12$^{+0.12}_{-0.14}$ & 0.07$^{+0.05}_{-0.05}$    & 0.08$^{+0.05}_{-0.05}$            & 0.26$^{+0.05}_{-0.10}$      & 0.24$^{+0.04}_{-0.07}$      
\\\\[-1em]
J2257-3306 & 2.59$^{+0.35}_{-0.40}$ & -0.16$^{+0.06}_{-0.06}$   & 0.37$^{+0.02}_{-0.04}$            & 0.21$^{+0.08}_{-0.16}$      & -0.03$^{+0.09}_{-0.09}$     
\\\\[-1em]
J2305-3436 & 2.28$^{+0.32}_{-0.35}$ & 0.32$^{+0.05}_{-0.06}$    & 0.10$^{+0.06}_{-0.05}$            & -0.23$^{+0.10}_{-0.05}$     & 0.25$^{+0.04}_{-0.07}$      
\\\\[-1em]
J2312-3438 & 1.30$^{+0.35}_{-0.34}$ & -0.10$^{+0.04}_{-0.04}$   & 0.11$^{+0.05}_{-0.05}$            & -0.10$^{+0.14}_{-0.08}$     & 0.20$^{+0.05}_{-0.05}$      
\\\\[-1em]
J2327-3312 & 0.88$^{+0.41}_{-0.34}$ & 0.05$^{+0.05}_{-0.06}$    & 0.04$^{+0.04}_{-0.03}$            & -0.18$^{+0.18}_{-0.12}$     & 0.18$^{+0.06}_{-0.05}$      
\\\\[-1em]
J2356-3332 & 1.24$^{+0.53}_{-0.48}$ & 0.09$^{+0.08}_{-0.08}$    & 0.32$^{+0.05}_{-0.07}$            & 0.09$^{+0.13}_{-0.19}$      & 0.13$^{+0.09}_{-0.10}$      
\\\\[-1em]
J2359-3320 & 2.55$^{+0.27}_{-0.30}$ & -0.07$^{+0.06}_{-0.06}$   & 0.32$^{+0.05}_{-0.07}$            & 0.14$^{+0.16}_{-0.20}$      & 0.25$^{+0.04}_{-0.07}$      \\ \bottomrule
\end{tabular}
\label{tab:stel_pop_results}
\end{table}

%\subsection{Correlations between the DoR and the stellar population parameters}
%\label{sec:DoR_vs_Stelpop}

\begin{figure*}
    \centering
    \includegraphics[width=\textwidth]{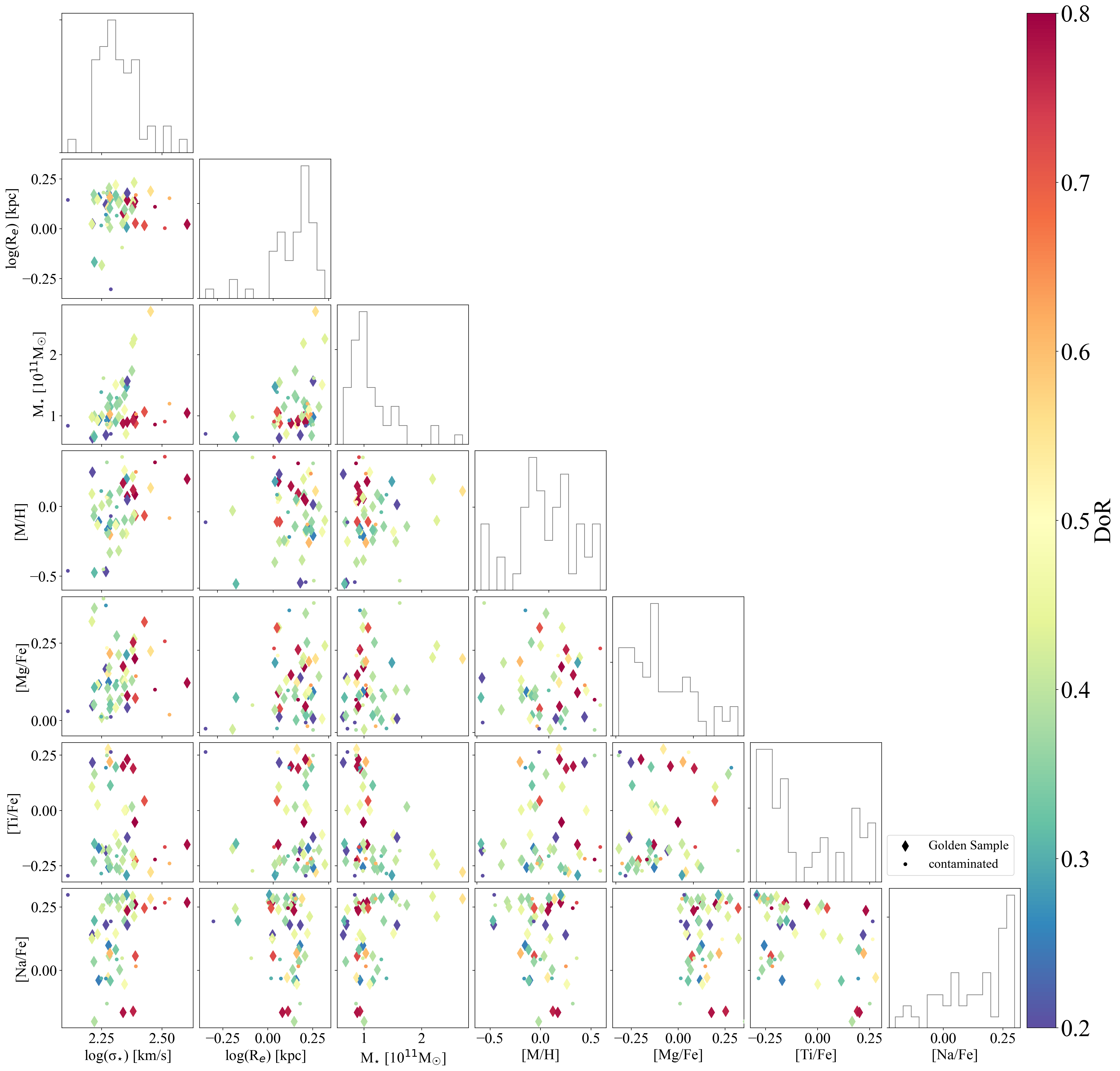}
    %{corner_plot.png}
    \caption{Stellar population analysis results. The data points are colour-coded by the DoR, computed in DR3. The only two parameters that show a correlation with it are the stellar metallicity and stellar velocity dispersion, which are both larger for relics than for non relics. Moreover, the most extreme relics (red points) seem to cluster  around $\log$(\Reff)$\,\sim0$ kpc and \Mstar$\sim1$-$1.5$ $\times10^{11}$M$_{\odot}$.}
    \label{fig:corner}
\end{figure*}

Figure~\ref{fig:corner} summarises the results of the FIF analysis and also shows how the stellar population parameters correlate with the DoR, the stellar mass, size and velocity dispersion of the \INSPIRE\ objects. In particular, the only two parameters that seem to correlate with the DoR are metallicity and velocity dispersion, as already \chiara{hinted for in \citetalias{Spiniello+23}}. More specifically, relics, and especially extreme relics have systematically larger stellar velocity dispersion than non-relics. 
For the metallicity, there is a large scatter but, all the extreme relics have super-solar [M/H] and a hint for a correlation can be seen. This will be further investigated in Section~\ref{sec:dor_metal}.
No correlation is found instead between the DoR and the considered elemental abundances (Mg, Na and Ti)\footnote{We note that in DR1 and DR3 we found a correlation between the DoR and the [Mg/Fe], estimated via line-indices. However, as already reported in \citet{Barbosa21}, the estimates obtained via different techniques do not match.}. \chiara{We caution the reader that the estimates for [Na/Fe], and especially [Ti/Fe],  could be biased  
both by the SNR of the data and by the uncertainties in the models. This is especially true for the TiO molecular bands, since the line lists feeding the response functions from the CvD models are incomplete, and these broad features also depend on other elements, such as Carbon. 
Nevertheless, we need to fit for them in order to break the IMF-elemental abundance degeneracy \citep[e.g.,][]{Spiniello15}}. 
%However, we note that galaxies with larger stellar masses have overall higher metallicities and [Na/Fe] but lower [Ti/Fe].

Another interesting result is that while a scatter exists in the sizes and masses of the entire sample, although they are selected to be similarly ultra-compact and massive, the most extreme relics cluster around \Reff$\sim1$ kpc and $\sim1$-$1.5$ $\times10^{11}$M$_{\odot}$. Nevertheless, we stress that only with higher spatial resolution data (e.g., from space or using AO-supported ground telescopes) we can obtain a more precise and robust estimate of the size of these objects.  %Finally, there is no dependency between the $\alpha$-, Na-, and Ti-abundances and the DoR. However, we note that galaxies with larger stellar masses have overall higher metallicities and [Na/Fe] but lower [Ti/Fe].

\subsection{Correlation between stellar metallicity and DoR}
\label{sec:dor_metal}
\begin{figure}
    \centering    \includegraphics[width=\linewidth]{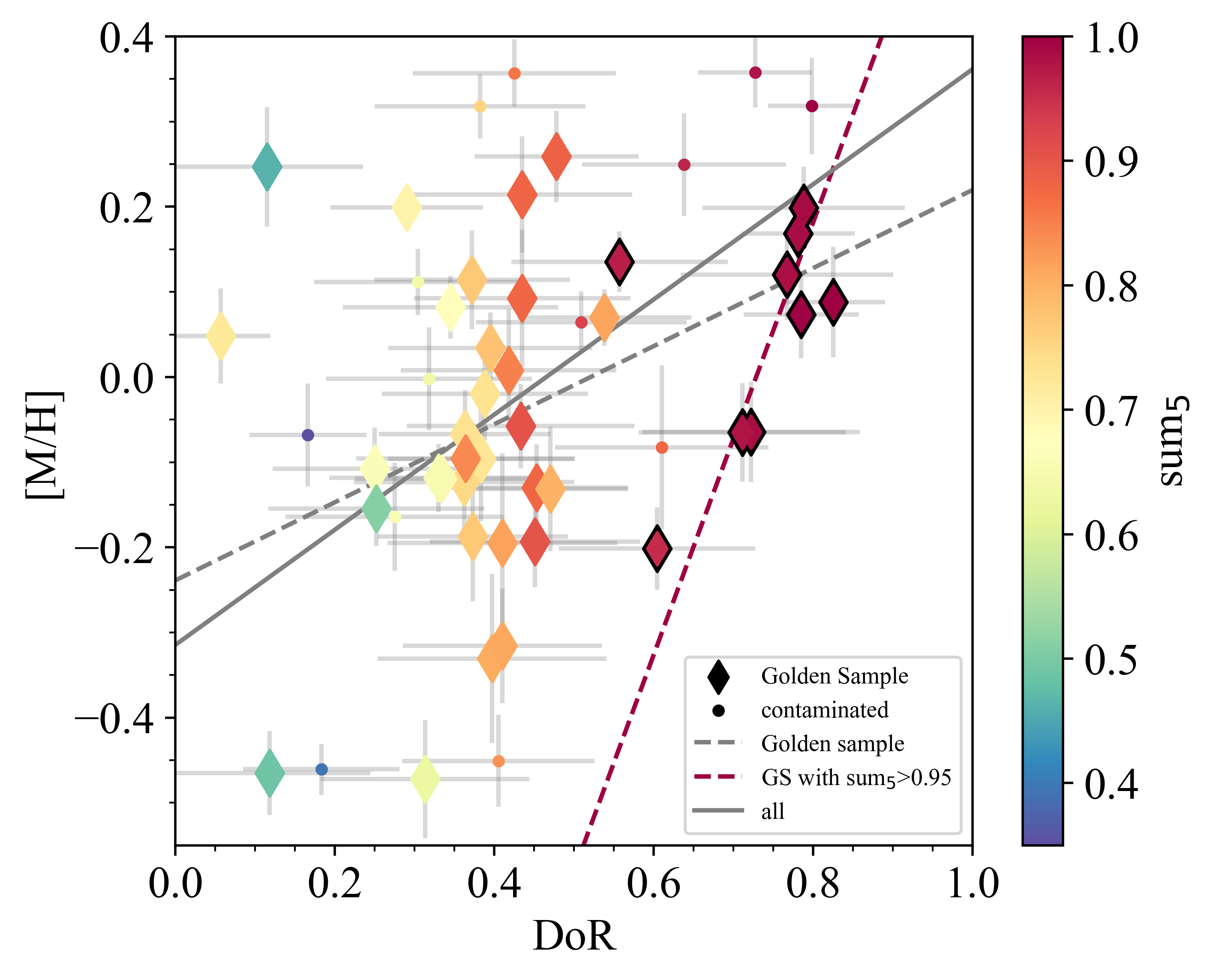}
    \caption{FIF fitted metallicity against the DoR, colour-coded by the maximum sum of weights of 5 neighbouring bins of the age-distribution, quantifying how good the single-stellar population assumption is. Diamonds represent galaxies from the \textit{Golden Sample}, while dots show the galaxies with contaminated spectral indices. \textit{Golden Sample} galaxies with sum$_{5}>0.95$ are marked by dark border. The lines represent the linear fits for the three samples. }\label{fig:dor_vs_metal}
\end{figure}

In Figure~\ref{fig:dor_vs_metal} we focus on the correlation between [M/H] and the DoR, colour-coding the data points by the number of bins defined above, to assess the validity of the SSP assumption.  

\begin{table}
\caption{Linear fits to the metallicity versus the DoR (Fig.~\ref{fig:dor_vs_metal})
for different \chiara{sub-samples} of galaxies. The last two columns report the Pearson correlation coefficient ($r$) and its p-value.}
\label{tab:linear_met}
\centering
\begin{tabular}{c|l|c|c}
\hline
Sample & Linear fit & $r$ & $p$ \\
\hline 
All & [M/H]$=[0.68\times \rm{DoR}]-0.315$ & $0.38$ & $0.005$ \\
\textit{Golden Sample} & [M/H]$=[0.46\times\rm{DoR}]-0.24$ & $0.3$ & $0.061$ \\
\textit{GS} with sum$_{5}$>0.95 & [M/H]$= [2.5\times \rm{DoR}]-1.8$ & $0.44$ & $0.24$ \\
\hline
\end{tabular}    
\end{table}

A very large spread in metallicity is observed for objects with a low DoR, which also corresponds to a larger number of different ages that \ppxf\ uses for the fit. 
Indeed, for objects that pass the threshold having sum$_{5}>0.95$, %extreme relics (DoR$\ge0.7$) 
the scatter is reduced and a clear direct correlation between the DoR and [M/H] is found. 
%Objects with a low DoR show a large spread in metallicity which is around solar and reaches at most a value of [M/H]$\sim0.2$ dex, while extreme relics (DoR$>0.7$) have always super-solar metallicities, reaching values as high as [M/H]$\sim0.4$ dex. 
This correlation seems to hold, although with larger scatter both considering all galaxies in the \textit{Golden Sample} or even all the \INSPIRE\ objects. 
We use a trimmed least squares linear fit to draw the correlation and show the resulting fits in  Figure~\ref{fig:dor_vs_metal}. We also list them in Table~\ref{tab:linear_met} for all the galaxies in the \textit{Golden Sample}, only these with sum$_{5}>0.95$, and all the \INSPIRE\ objects. 
Finally, we compute, for each fit, the Pearson coefficient and its p-value to assess how probable a correlation between the stellar metallicity and the DoR is. 
As expected, the strongest correlation is found for the sub-group of relics with sum$_{5}>0.95$, for which the totality of the mass was formed during the first phase of the formation scenario. 

%\begin{itemize}
%\centering
%  \item[] Golden sample: [M/H]$=(0.46\times\rm{DoR})-0.239$, $r=0.30$, $p=0.06$, \\
%  \item[] GS with sum$_{5}$>0.95: [M/H]$= (2.5\times \rm{DoR})-1.8$, $r=0.44$, $p$=0.24, \\
%  \item[] all:\,\,\,\,\,\,\,\, [M/H]$=(0.68\times \rm{DoR})-0.315$, $r=0.38$, $p=0.0049$
%\end{itemize}

%Even including the galaxies outside the \textit{Golden Sample}, the linear relation holds and is still statistically relevant. 

%is instead plotted in Figure~\ref{fig:dor_vs_params}.Here, no correlation is found for any of the abundances, showing a large scatter in all panels of Fig~\ref{fig:dor_vs_params}.  %for the [Ti/Fe] and the [Na/Fe], while a hint for a direct relation could be seen for the \textit{Golden Sample} in the [$\alpha$/Fe]-DoR plot. In fact, there are no extreme relics with solar $\alpha$ while none of non-relics (DoR$<0.34$) are extremely $\alpha$ enhanced. However the scatter in the relation (if any) is much larger in this case. 

\subsection{Correlation between the IMF slope and the DoR}
\label{sec:IMF}
\begin{figure}
    \centering    \includegraphics[width=\linewidth]{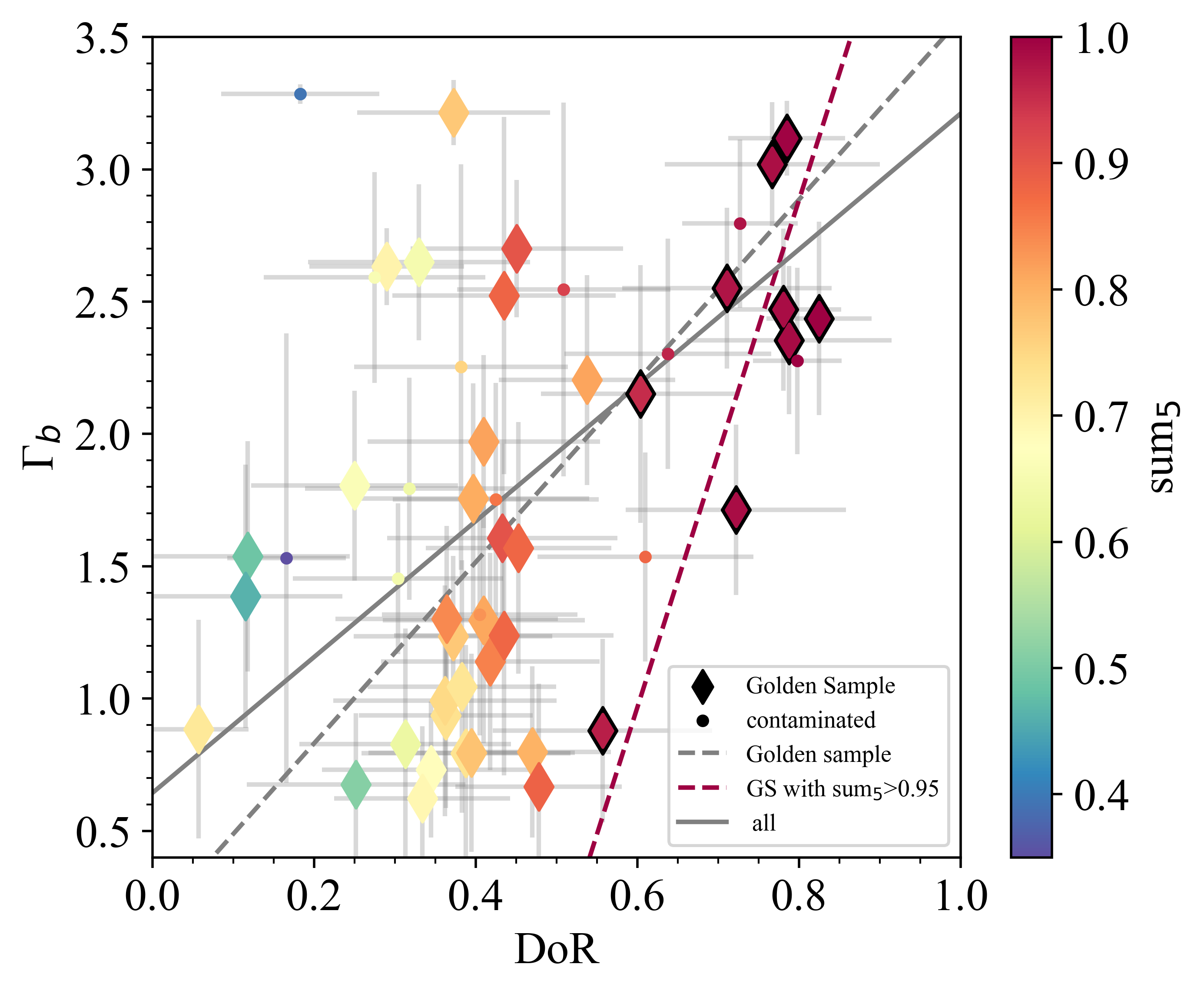}
    \caption{The slope of the IMF against the DoR, colour-coded by the maximum sum of weights of 5 neighbouring bins of the age-distribution. Symbols and lines are as in previous figures.}
\label{fig:imf_vs_dor}
\end{figure}

In this section, we focus on the stellar IMF and the correlation between its low-mass end slope and the DoR, which is shown in Figure~\ref{fig:imf_vs_dor}. 
Although with a non-negligible scatter, objects with a higher DoR are better fitted with a dwarf-richer IMF. 
Also in this case, we apply a least trimmed squares linear fitting to the data and report the results in Table~\ref{tab:linear_imf}. 

\begin{table}
\caption{Linear fits to the IMF slope versus the DoR (Fig~\ref{fig:imf_vs_dor}) for different sub-samples of galaxies. The last two columns report the Pearson correlation coefficient ($r$) and its p-value.}
\label{tab:linear_imf}
\centering
\begin{tabular}{c|l|c|c}
\hline
Sample & Linear fit & $r$ & $p$ \\
\hline 
All & $\Gamma_{b}=[2.57\times \rm{DoR}]+0.64$ & $0.4$ & $0.0033$\\
\textit{Golden Sample} & $\Gamma_{b}=[3.42\times \rm{DoR}]+0.15$ & $0.5$ & $0.0012$ \\
\textit{GS} with sum$_{5}$>0.95 & $\Gamma_{b}=[9.6\times \rm{DoR}]-4.8$ & $0.73$ & $0.024$ \\
\hline    
\end{tabular}    
\end{table}

The strongest correlation is found for the \textit{Golden Sample} galaxies with a very peaked age distribution (sum$_{5}>0.95$), however, a significant correlation is found also when considering all the galaxies. 

All the extreme relics, i.e. DoR$>0.7$, are inconsistent with having a Milky-Way IMF and require a steeper slope. However, we note that there are objects with a DoR as low as $\sim 0.3$ that fit with an IMF slope equally steep than objects with a higher DoR. 
%The correlation coefficient is 0.47 with a p-value of 0.001, hence implying a correlation between these two quantities. 
This result is fully consistent with the results in \citetalias{Martin-Navarro+23} in which we found a larger scatter in the IMF slope of non-relics than that of relics, 
which might be caused by the fact that the SSP assumption does not hold well for galaxies with a more extended SFH, that could have stellar populations with different IMFs (and age and [M/H]). 
\chiara{Furthermore, it is important to stress that data and modelling systematics might play a role, increasing the scatter. Indeed, although we carefully checked each individual spectrum, we are aware that the SNR is, in some cases, towards the lower limit for which FIF produces trustable results. }

\begin{figure*}
    \centering
    \includegraphics[width=\textwidth]{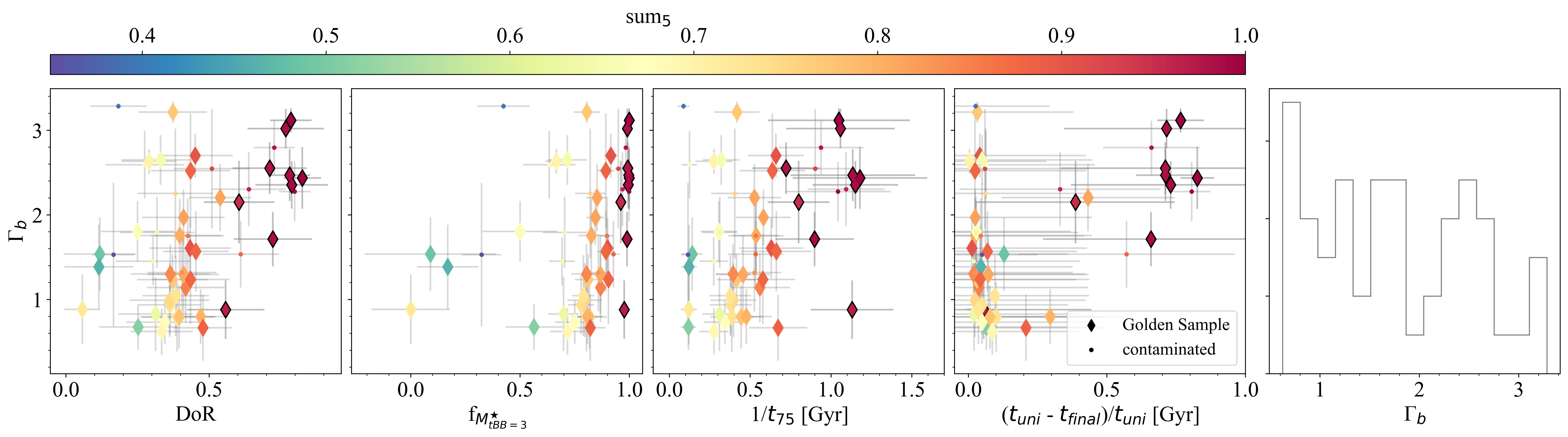}
    \caption{The slope of the IMF against the DoR (left) and the three quantities that were used to define it, \chiara{and the distribution of $\Gamma_{b}$ inferred for the 52 UCMGs (histogram in the rightmost  panel)}. The datapoints are colour-coded by the maximum sum of weights help by 5 neighbouring bins of the age-distribution. Diamonds represent galaxies from the \emph{Golden sample}, while dots mark the galaxies with contaminated spectral indices.}
\label{fig:imf_vs_defs}
\end{figure*}

To better understand why some galaxies with a relatively low DoR have a bottom-heavy IMF, we have considered each of the three addenda that went into the definition of the DoR (from \citetalias{Spiniello+23}) separately.  We remind the readers that the DoR is defined as the sum of 
the fraction of stellar mass assembled at $z=2$ (i.e., \Mfrac), the inverse of the cosmic time at which 75\% of the stellar mass was in place ($t_{75}$), and the inverse of the final assembly time (i.e. the time at which a galaxy has formed the 100\% of its stellar mass), re-scaled to the age of the Universe at the redshift of each galaxy ($[t_{\text{uni}}-t_{\text{fin}}]/t_{\text{uni}}$). 
The relation between the IMF slope and each of these three quantities is plotted in Figure~\ref{fig:imf_vs_defs}, with the same symbology used in other figures. \chiara{Interestingly, the strongest relation is found with the \Mfrac: there are no points with $\Gamma_b>2$ for \Mfrac$<0.6$. Hence, %if 
UCMGs have a dwarf-rich IMF  only when they have formed at least 60\% of their stellar mass at $z>2$}. %, ii) they have formed more than 75\% of their stellar mass 4 Gyr after the BB, even if they then continued to form stars until their current redshift. 

%We remind the reader that the DoR has been defined as an arbitrary dimentionless number varying from 0 to 1. 
In conclusion, there is a clear dependency of the IMF slope from the cosmic assembly time \chiara{(second-left panel of Figure~\ref{fig:imf_vs_defs})}, \chiara{especially for objects with a very peaked SFH and very old ages}. The IMF is dwarf-richer than the Milky-Way one for stellar populations formed at $z>2$, i.e., during the first phase of the two-phase formation scenario.

%\begin{figure*}
%\centering
%\includegraphics[width=\linewidth]{imf_vs_defs.png}
%\caption{The slope of the IMF against the three quantities that were used to define the degree of relicness, colour-coded by DoR. Crosses represent galaxies from the \emph{Golden sample}, while dots mark the galaxies with contaminated spectral indices. \textcolor{red}{NECESSARY TO CHANGE THE RIGHTMOST PANEL: We need to "rescale" for tuni.}}   
%    \label{fig:imf_vs_other}
%\end{figure*}

\begin{figure*}
    \centering
\includegraphics[width=0.8\linewidth]{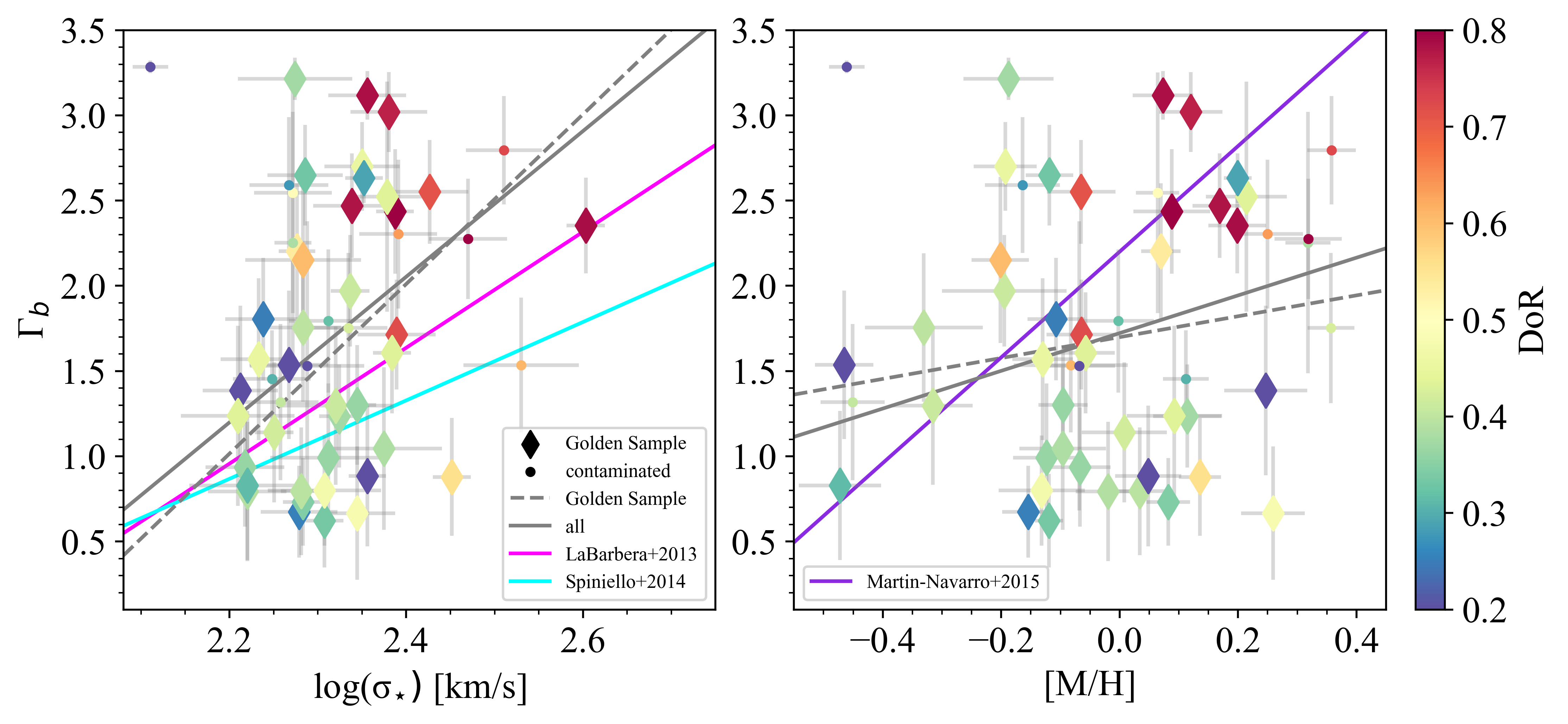}
%{result_plot_3.png}
\caption{Stellar IMF slope versus stellar velocity dispersion (left, computed in DR3) and total stellar metallicity (right, computed with FIF). The \INSPIRE\ galaxies are colour-coded by their degree of relicness (top row) and sum$_{5}$ (bottom row). \chiara{The lines show linear fits to the data, from this work (grey) and from literature papers, as reported in the legend and in the third and fourth block of Table~\ref{tab:linear_sigma_metal}. Symbols are like in other figures.}} %Diamonds represent galaxies from the \textit{Golden sample}, while dots mark the galaxies with contaminated spectral indices.
\label{fig:imf_vs_veldisp}
\end{figure*}

\subsection{Correlation between the IMF slope and other parameters}
\label{sec:literature}
\chiara{Two of the most accredited relations involving global IMF variation are the IMF slope-velocity dispersion \citep{Cappellari+12, Cappellari+13_ATLAS3D_XX, Ferreras+13, LaBarbera+13_SPIDERVIII_IMF, Spiniello+14} and the IMF-metallicity relation \citep{Martin-Navarro15_CALIFA}. 
In Figure~\ref{fig:imf_vs_veldisp}, we check whether we are able to reproduce them with the \INSPIRE\ UCMGs. 
The least-square linear fits to the data, together with the Pearson coefficient ($r$) and its p-value, are reported in Table~\ref{tab:linear_sigma_metal}.

A strong correlation is found with the $\sigma_{\star}$, as shown by the least square fitting to the data (left panel, grey lines), both all the galaxies (solid) and only these belonging to the \textit{Golden Sample} (dashed). 
The linear relation is even steeper than that found from the literature on normal-sized ETGs covering a similar range in $\sigma_{\star}$ (\citealt{LaBarbera+13_SPIDERVIII_IMF}, magenta line, and \citealt{Spiniello+14}, cyan line). However, we also note that for the same value of $\sigma_{\star}$, we find a very large range of IMF slopes, but that objects with a higher DoR have systematically steeper IMF slopes. %This is not true however when considering the sum$_{5}$ (bottom panel). In addition, when fitting only objects with sum$_{5}>0.95$, the slope of the $\Gamma_{b}$-$\log{\sigma_{\star}}$ relation changes dramatically. 

The correlation of the IMF slope with the [M/H] is instead much weaker than the one reported in \citet{Martin-Navarro15_CALIFA}.  Furthermore, also in this case, we note that at equal metallicity,  
%Moreover, generally, at equal velocity dispersion and metallicity, 
galaxies with a higher DoR have a dwarf-richer IMF. 
%In this case, a similar trend with however a systematical shift towards higher metallicity is found when considering only UCMGs with sum$_{5}>0.95$.  

Hence we conclude that even though clear and solid global relations have been reported \citep{LaBarbera+13_SPIDERVIII_IMF, Spiniello+14, McDermid+14_IMF, Martin-Navarro15_CALIFA}, which seem to hold also for UCMGs, there might not be a casual connection between the IMF slope and the stellar velocity dispersion and metallicity. Or, at least, these two factors are not the only drivers of the variations in the low-mass end of the IMF. 
Hints from the fact that the local value of the stellar velocity dispersion is not the main driver behind dwarf-to-giant ratio variations was already suggested by \citet{Martin-Navarro+15_IMF_relic} for the local relic NGC1277.

We speculate that the relations might arise from the fact that by selecting a galaxy with a larger velocity dispersion (i.e. a more massive galaxy), or with a metal richer population, the chances to find a compact progenitor in its centre (or a relic) are higher \citep{Pulsoni21}. And that the low-mass end of the IMF slope might be steeper for it. 
This is for instance the case of NGC~3311, the central galaxy of the Hydra-I cluster. From very high SNR spatially resolved spectra, \citet{Barbosa21} found a bottom-heavy IMF slope in the galaxy's centre, where the stellar velocity dispersion was as low as $\sim150$ \kms. In this region, the stars are incredibly old and metal-rich. The IMF becomes instead Milky-Way-like moving away from the centre, where the $\sigma_{\star}$ rises up to $\sim 400$ \kms, and both age and metallicity drop.  }

\begin{table}
\caption{Linear fits to the IMF slope versus the stellar velocity dispersion (first block) and the metallicity (second block) 
for different sub-samples of galaxies. The lines are shown in Figure~\ref{fig:imf_vs_veldisp}. The last two columns report the Pearson correlation coefficient ($r$) and its p-value.}
\label{tab:linear_sigma_metal}
\centering
\begin{tabular}{c|l|c|c}
\hline
Sample & Linear fit & $r$ & $p$ \\
\hline 
\textit{Golden Sample} & $\Gamma_{b}=[5 \times \rm{log(\sigma_{\star})}]-9.9$ & $0.33$ & $0.043$ \\
All & $\Gamma_{b}=[4.3 \times \rm{log(\sigma_{\star})}]-8.2$ & $0.19$ & $0.18$\\
\hline 
\textit{Golden Sample} & $\Gamma_{b}=[0.61\times $[M/H]$]+1.7$ & $0.12$ & $0.45$ \\
All & $\Gamma_{b}=[1.11\times $[M/H]$]+1.72$ & $0.13$ & $0.35$\\
\hline
\end{tabular}    
\end{table}

\section{Discussion}
\label{sec:discussion}
The results presented in this paper unambiguously demonstrate that a correlation exists between the low-mass end of the IMF, the stellar metallicity and the cosmic epoch of star formation. In particular, we found that, among the 52 \INSPIRE\ UCMGs,  the objects with the highest DoR, i.e. those that formed the totality of their stellar masses at $z>2$ and did not experience any other star formation episode, \chiara{i.e. extreme relics}, are metal richer and have a dwarf-rich IMF. 

%In particular, the \INSPIRE\ data have shown that objects that have assembled the majority of their stellar mass early on in cosmic time and very quickly, tend to be metal richer and have a dwarf-rich IMF. 
This result could seem in strong disagreement with the latest findings from JWST data, suggesting that the IMF might be top-heavy, i.e.~giant richer, at very high redshift \citep{Cameron23, Trinca23, Woodrum23}. 
\chiara{However, we note that, in principle, the high- and low-mass ends of the IMF are not necessarily coupled. 
Here below we highlight a couple of possible scenarios that would allow us to reconcile the different observations. }

In the context of the integrated galaxy-wide stellar initial mass functions (IGIMFs) paradigm \citep{Kroupa95, Kroupa03}, 
all stars form in groups or embedded star clusters, and the most massive star that can be formed in them depends on the mass and the metallicity of the star cluster \citep{Weidner10, Dabringhausen23}. The IGIMF will then be the combination of all the stars that 
formed in the different star clusters of a galaxy \citep{Kroupa03}. 
Furthermore, recent simulations on the first, metal-free stars suggest that the fraction of massive stars directly depends on the gas temperature of star-forming clouds: the higher the temperature the larger the mass of the produced stars \citep{Abel02, Fukushima20}.  
Also, the top-heaviness of the IMF depends on the metallicity of the gas (the lower the metallicity the more top-heavy the IMF would be, \citealt{Fukushima20}) and on the background radiation intensity (the higher the intensity the more top-heavy the IMF would be, \citealt{Chon22}). 
Hence, one possibility is that the IMF could be top-heavy in low metallicity and dense gas environments \citep{Dabringhausen+12, Dib23} while it becomes progressively bottom-heavy as the metallicity of the environment increases \citep{Chabrier_2014}, at any redshifts. \chiara{However, we do not find a strong dependency of the low-mass end of the IMF on the measured stellar metallicity. } %According to this scenario, then, the high-$z$ Lyman-$\alpha$-emitting galaxies would not be the progenitors of the relics and the cores of massive ETGs. 

Another interesting possibility, instead, is to allow for a time-evolution of the IMF slope \citep{Vazdekis96, Vazdekis97, Weidner+13_giant_ell, Ferreras15}. 
A first and quick phase with a top-heavy IMF occurs at a very high redshift ($z\ge5$). Then, the very massive ($M>50M_{\odot}$) giant stars rapidly die (less than 10 Myr, \citealt{Yusof13}), polluting the interstellar medium with metals. At this point, if the conditions of the local environment (i.e., gas temperature, pressure and density) are extreme, and a violent starburst happens, fragmentation of the star-formation clouds becomes easier \citep{Chabrier_2014} and therefore a large number of dwarf stars are produced, transforming the IMF into a bottom-heavy one. In fact, according to theoretical works (e.g.~\citealt{Hennebelle08}), a larger fraction of low-mass cores in very dense, hot, and highly turbulent environments is a direct consequence of the enhanced gas compression by highly turbulent motions and of the shorter free-fall times for the collapsing over-dense region \citep{Chabrier_2014}. 
\chiara{This scenario also naturally explains why relics are metal-richer than non-relics, as they formed in a metal-rich gas. }
Finally, stars are produced following a canonical Milky-Way-like IMF slope for the more time-extended standard star formation occurring during the second phase of the two-phase formation scenario. 

\section{Conclusions}
\label{sec:conclusion}
In this paper, the sixth of the INvestigating Stellar Populations In RElics (INSPIRE, \citealt{Spiniello20_Pilot, Spiniello+21}) series, we have presented an extensive stellar populations analysis on the entire \INSPIRE\ sample made of 52 ultra-compact (\Reff$<2$ kpc) massive ($M_{\star}>6\times 10^{10}$ $M_{\odot}$) galaxies (UCMGs) at $0.1<z<0.4$. Of these, 38 were confirmed as relics in \citetalias{Spiniello+23}, as they formed more than 75\% of their stellar mass by the end of the first phase of the formation scenario. At the time of writing, this is the largest catalogue of spectroscopically confirmed relics known in the nearby Universe. 

In \citetalias{Spiniello+23} the stellar population ages and metallicities were inferred fitting UVB+VIS spectra, joined and convolved to a common final resolution of FWHM = 2.51\AA, in the wavelength range [3500-7000]\AA. The [Mg/Fe] abundances were instead obtained directly from line index analysis, with the Mg$_b$-Fe24 index--index plot. The IMF slope was kept fixed to a Kroupa-like one with $\Gamma_b = 1.3$ and all other elemental abundances were assumed to be solar. Here, instead, we have used the Full-Index Fitting (FIF) technique to infer stellar metallicities,  [Mg/Fe], [Ti/Fe] and [Na/Fe] ratios, and IMF slope from the same spectra.
This much more flexible approach, combined with a careful visual inspection of each single spectrum, allowed us to investigate whether a correlation exists between the stellar population parameters and the degree of relicness. \chiara{We have been also able,  for the first time, to infer the low-mass end of the IMF slope on individual galaxies}. Moreover, we have tested for which systems the SSP assumption holds, by quantifying the spread in age between all the stellar models used by the \ppxf fit \chiara{and assesses whether the derived relation depends upon this quantity}. 
\smallskip

In particular, we found: 
\begin{itemize}
    \item a clear correlation between the velocity dispersion and the DoR, confirming the results already presented in previous \INSPIRE\ papers: at equal stellar mass relics have larger $\sigma_{\star}$ than non-relics;
    \item a linear relation between the stellar metallicity and the DoR, although with a large scatter. The slope of the relation is much steeper considering only extreme relics for which the SSP assumption certainly holds;
    \item a correlation between the IMF slope and the DoR, with a scatter that increases when the stellar population fit requires a larger number of SSP models with different ages. Specifically, all UCMGs with DoR$>0.7$, i.e. extreme relics,  require an IMF which is dwarf-richer than that of the Milky Way, with slopes of $\Gamma_b\ge1.8$. For UCMGs with lower DoR, the spread in the IMF slope becomes very \chiara{large}, consistent with what was found in \citetalias{Martin-Navarro+23}. 
    \item \chiara{considering the three components forming the DoR, there are no UCMGs with $\Gamma_{b}>2$ when the fraction of stellar mass formed by $z\sim2$ is lower than 60\%}; 
    %We therefore conclude that the IMF slope might be 
%This is because for systems with intermediate (and low) DoR the SSP assumption is not as solid as for extreme relics. Indeed, for these systems, not all the mass has been formed during the first phase of the formation scenario. If the IMF bottom-heavy only in stars formed early-on in cosmic time, under extreme conditions \citep{Chabrier_2014}}; 
    \item \chiara{a correlation between the slope of the IMF and the stellar velocity dispersion, which is even steeper than the ones from the literature;  
    \item a very weak correlation (or no correlation) between the IMF slope and the stellar metallicity; 
    \item that at equal velocity dispersion and/or metallicity, galaxies with a higher DoR have a dwarf-richer IMF; }
    \item no correlation between the DoR and the other stellar population parameters we infer ([Mg/Fe], [Na/Fe], [Ti/Fe]). %which, however, correlate with stellar mass: more massive systems have larger [Na/Fe] but lower [Ti/Fe]; 

%    \item comparing UCMGs with normal-sized ETGs of similar dynamical masses,  ...\textcolor{red}{TO BE CONTINUED}

\end{itemize}
%We found linear correlations between the stellar metallicity and the DoR and between the stellar IMF slope and the DoR. 
%Objects with a higher DoR, i.e. objects that formed through a more extreme SF event at higher-$z$ are metal richer (Eq.~1) and have a steeper IMF slope (Eq.~2). 
In conclusion, our data supports a scenario whereby 
an excess of dwarf stars might originate from the first phase of the mass assembly at high-z when the density and temperature of the Universe were higher, and thus when fragmentation might have been easier \citep{Chabrier_2014}.  However, to reconcile the very recent findings at very high-z from JWST, we speculate that the IMF might vary with cosmic time \citep{Vazdekis96, Vazdekis97, Weidner+13_giant_ell, Ferreras15}. \chiara{An incredibly quick top-heavy phase at very high-z ($>5$), that produces very metal-poor stars of $\sim100$ M$_{\odot}$. These stars quickly die and pollute the interstellar medium with metals. Then, a large number of dwarfs is produced in regions of high density and temperature, through fast starbursts up to $z\sim2$, while stars are distributed with a Milky-Way-like IMF slope if they formed in the second phase of the two-phase formation scenario, under less extreme temperature and density conditions. }

%%%%%%%%%%%%%%%%%%%%%%%%%%%%%%%%%%%%%%%%%%%%%%%%%%
\section*{Data Availability}
The data described in this paper are publicly available via the ESO Phase 3 
Archive Science Portal under the collection  \INSPIRE\ (\url{https://archive.eso.org/scienceportal/home?data_collection=INSPIRE}).

\section*{Acknowledgements}
The authors are thankful to Prof. Vazdekis for interesting discussions the improved the quality of the paper. CS, CT, DB and PS acknowledge funding from the PRIN-INAF 2020 program 1.05.01.85.11. 
AFM acknowledges support from RYC2021-031099-I and PID2021-123313NA-I00 of MICIN/AEI/10.13039/501100011033/FEDER,UE,NextGenerationEU/PRT. 
DS is supported by JPL, which is operated under a contract by Caltech for NASA.



\bibliographystyle{mnras}
\bibliography{biblio_INSPIRE}

%%%%%%%%%%%%%%%%%%%%%%%%%%%%%%%%%%%%%%%%%%%%%%%%%%

%%%%%%%%%%%%%%%%% APPENDICES %%%%%%%%%%%%%%%%%%%%%

\appendix

\section{Golden Sample selection}
\label{app:testing}

\begin{figure*}
\centering
  \centering
  \includegraphics[width=0.8\linewidth]{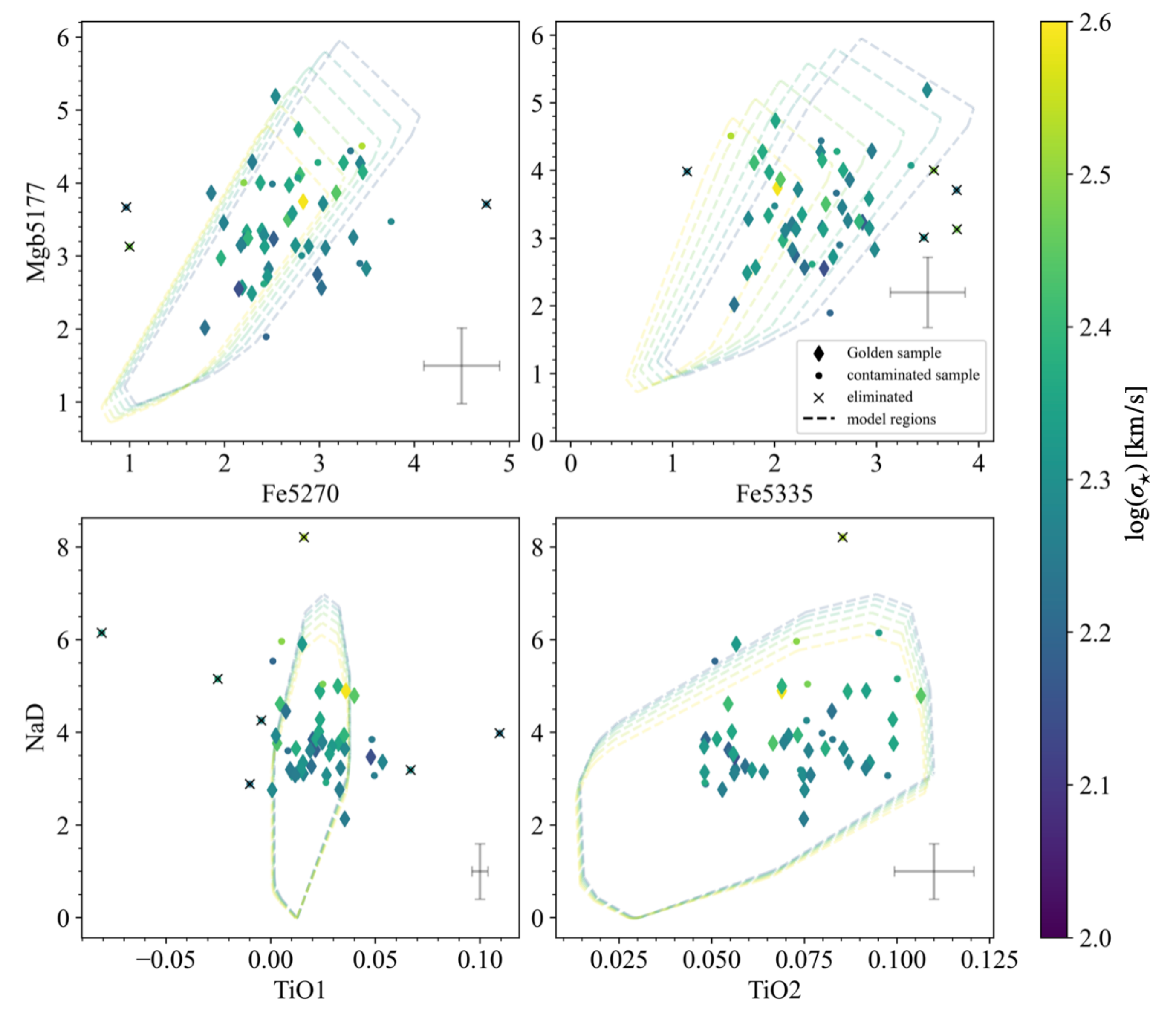}
  \caption{ Examples of four index-index plots showing the SSP model grid and the measurements for the \INSPIRE\ galaxies colour-coded by their resolution/velocity dispersion. Systems that have no contaminated indices (\textit{Golden sample}) are marked by crosses, while systems that have at least one contaminated index are marked by a dot. The black circles mark the systems that were eliminated on each panel based on the corresponding indices. A typical error-bar is reported in the bottom right corner of each panel.}
\label{fig:ind-ind_plot}
\end{figure*}

\begin{table*}
\centering
\begin{tabular}{cccccccc}
\hline \\[-1em]

\textbf{Run} & \textbf{\begin{tabular}[c]{@{}c@{}}Hydrogen \\ index \end{tabular}} & \textbf{\begin{tabular}[c]{@{}c@{}}Magnesium\\  index \end{tabular}} & \textbf{\begin{tabular}[c]{@{}c@{}}Iron\\  index\end{tabular}} & \textbf{\begin{tabular}[c]{@{}c@{}}Titanium\\  index\end{tabular}} & \textbf{\begin{tabular}[c]{@{}c@{}}Sodium\\  index\end{tabular}} & \textbf{Notes}  & \textbf{\begin{tabular}[c]{@{}c@{}} \# of \\ galaxies \end{tabular}} \\\\[-1em] \hline\\[-1em]

fiducial &    -                   & Mgb5177     & Fe5270, Fe5335                 & TiO1, TiO2       & NaD &         & 39 \\
\hline\\[-1em]
1  &    -                  & Mgb5177     & Fe5270, Fe5335                 & TiO1, TiO2       & -       & not fitting for {[}Na/Fe{]} & 40 \\
2 & -                     & Mg2 & Fe5270, Fe5335                 & TiO1, TiO2       & NaD &     testing a different Mg index    & 39 \\
3 & -                     & Mgb5177 & Fe5270, Fe5335                 & TiO1, TiO2       & - &         & 40 \\
4  & -                     & Mgb5177     & Fe5270, Fe5335                 & TiO1, TiO2, TiO3 & NaD &   testing more TiO lines      & 35 \\
5  & -                     & Mgb5177     & Fe5015, Fe5270, Fe5335, Fe5709 & TiO1, TiO2       & NaD &     testing more Fe lines    & 31 \\
6 & -                     & Mgb2 & Fe5015, Fe5270, Fe5335, Fe5709 & TiO1, TiO2       & NaD &         & 31 \\
\\[-1em]\hline\\[-1em]
7 & H$\beta$G             & Mgb5177 & Fe5270, Fe5335                 & TiO1, TiO2       & NaD & age fixed        & 38 \\
8  & H$\beta$G             & Mg2     & Fe5270, Fe5335                 & TiO1, TiO2       & NaD &     age fixed, different Mg index    & 38 \\
9 & H$\beta$G              & Mg2     & Fe5270, Fe5335                 & TiO1, TiO2       & NaD & age free to vary & 38 \\
10  & H$\delta$A            & Mg2     & Fe5270, Fe5335                 & TiO1, TiO2       & NaD &      testing a different age index   & 37 \\
11 & H$\delta$A            & Mgb5177 & Fe5270, Fe5335                 & TiO1, TiO2       & NaD &         & 37 \\
12 & H$\beta$G, H$\delta$A & Mgb5177 & Fe5270, Fe5335                 & TiO1, TiO2       & NaD &         & 36 \\
13  & H$\beta$G, H$\delta$A & Mg2     & Fe5270, Fe5335                 & TiO1, TiO2       & NaD &       & 36 \\
\\[-1em]\hline\\[-1em]
14 & H$\beta$G             & Mgb5177 & Fe5015, Fe5270, Fe5335, Fe5709 & TiO1, TiO2       & NaD &   adding more iron lines      & 29 \\
15  & H$\beta$G             & Mg2     & Fe5015, Fe5270, Fe5335, Fe5709 & TiO1, TiO2       & NaD &         & 29 \\
16 & H$\delta$A            & Mgb5177 & Fe5015, Fe5270, Fe5335, Fe5709 & TiO1, TiO2       & NaD &         & 30 \\ 
17 & H$\delta$A            & Mg2     & Fe5015, Fe5270, Fe5335, Fe5709 & TiO1, TiO2       & NaD &         & 30 \\
18 & H$\beta$G, H$\delta$A & Mgb5177 & Fe5015, Fe5270, Fe5335, Fe5709 & TiO1, TiO2       & NaD &         & 28 \\
19  & H$\beta$G, H$\delta$A & Mg2     & Fe5015, Fe5270, Fe5335, Fe5709 & TiO1, TiO2       & NaD &         & 28 \\
\\[-1em]\hline
\end{tabular}
\caption{FIF runs with different configurations. The first row reports our fiducial run, based on 6 indicators: Mgb, Fe5270, Fe5335, NaD, TiO1, and TiO2. The other 18 runs, have been used to test different assumptions, indices and setups, as described in details in the text. }
\label{tab:runs}
\end{table*}

In Section~\ref{sec:FIF}, we have selected six indices that have allowed us to infer the age, metallicity, [Mg/Fe], [Ti/Fe], [Na/Fe] abundances, and IMF slope.  
To assess which of these features were contaminated in each of the galaxies we have performed three different tests:

\begin{itemize}
    \item[i)] we have visually inspected all the 52 spectra, flagging clear emission lines and residuals contaminating the indices,
    \item[ii)] we have plotted the histogram distribution of the indices computed on all the models varying all the parameters, and have flagged systems for which a given index falls outside this distribution,
    \item[iii)] we have produced as many index-index plots as possible, creating large model grids with changing resolution (in \kms) and elemental abundance.
\end{itemize}

The result of this latter test is visualised in Figure~\ref{fig:ind-ind_plot}. Here we show four representative index-index plots (top: Mg2 versus the two Fe lines, bottom: NaD vs the two TiO lines) and a series of grids predicted by the MILES SSP models. 
For each panel, if a galaxy falls outside the model grid, it is highlighted with a black circle and flagged as 'contaminated'. 
We note that we have performed these tests considering a larger set of indices but still arising from the same chemical elements (H$\beta$G, H$\beta$, H$\beta$0, Mg2, Mgb5117, Fe5015, Fe5270, Fe5335, Fe5709, NaD, H$\delta$A, TiO1, TiO2, TiO3, TiO4). 
This was done in order to select, for each element the least-contaminated index. 

Indeed, 
inspecting by eye each of the spectra, we found that for 40/52 UCMGs, the six indices mentioned above are clean from residual sky-lines and bad pixels.
From these, however, we exclude two more objects that have a non-negligible percentage ($\ge25\%$) of stars younger than 1 Gyrs. This is because in these two cases, the assumption of an SSP model is not a valid one. 
The final sample of 39 uncontaminated, old galaxies \chiara{has thus been} denoted as the {\sl Golden Sample}. All the plots presented in the main body of the paper always show them as diamonds, while the objects for which one or more indices are contaminated are plotted as circles.  

%We note that we only use a (large) combination of SSPs. This is a very good approximation for massive galaxies with old stellar populations (e.g. \citealt{Thomas+05}) but poses some problems when fitting galaxies with a very extended star formation history (SFH) and with populations younger than 1-2 Gyr. Hence, a priory, we eliminated from the \INSPIRE\ sample we consider for this analthe two non-relic galaxies: J1142+0012 and  J2327-3312.  They both have $\ge25$\% of their stars that formed $\le 1$ Gyr ago. 

At this point, as a further test for the robustness of our results, we have made several modelling runs with different combinations of spectral indices and different setups (e.g., not fitting for [Na/Fe] or/and not fixing for the age and hence including or not Balmer lines).  
All the runs are summarised in Table~\ref{tab:runs}. 

\begin{figure*}
\centering
\includegraphics[width=\linewidth]{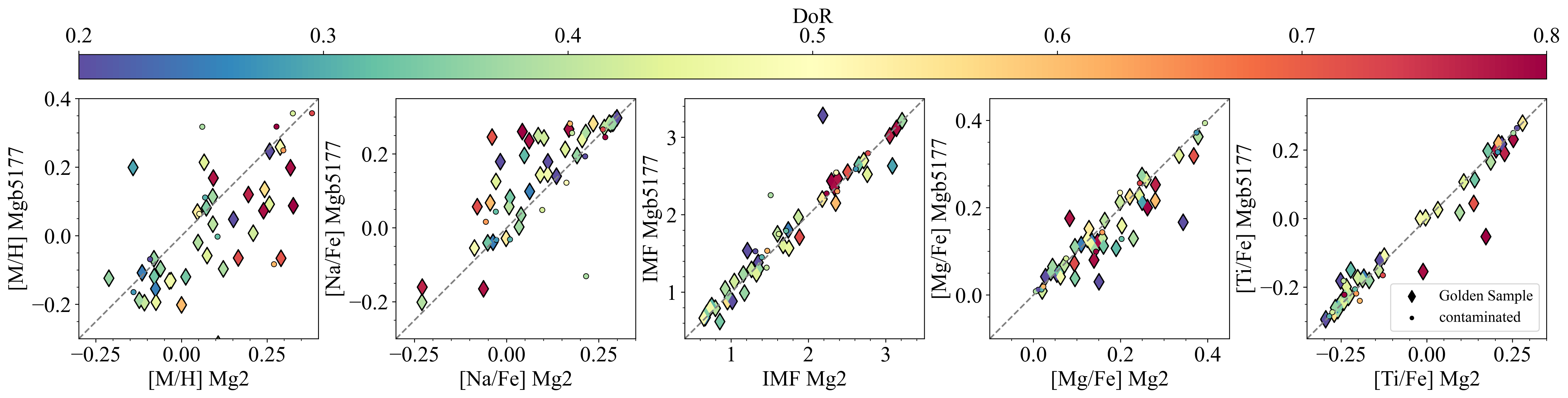}
\caption{Comparison between the stellar population parameters obatained using the two different spectral index definitions for Magnesium (Mg2 and Mg5177). The data points are colour-coded by the DoR. }   
    \label{fig:mg_comparison}
\end{figure*}

Firstly (Runs 1-6), we investigated whether the results would differ greatly by excluding/including other spectral indices from the fiducial ones. In particular, in Run~1 and 3 we have excluded the NaD index and not fitted for Na abundance, which did not lead to any notable changes in the modelled values. We note that we do not test the case without TiO indices and fixing the [Ti/Fe] to solar because the TiO indices are the best gravity-sensitive indicators and we need to include them in order to have a solid constraint on the IMF slope, which is the main goal of this paper. 
In Run~2 we have used an alternative definition of the Mgb spectral feature (Mg2 instead of the more classical Mgb5177, see Table~\ref{tab:indices} for the index definition), which led to larger metallicities at the cost of lower Na abundance, as seen in Figure \ref{fig:mg_comparison}. These changes are systematic but are smaller for more extreme relics. Importantly, the IMF slopes inferred from the two runs are consistent within the errors, demonstrating that the inference on this parameter is robust. 
%We note that the metallicities we would infer with Mgb would be too low given the stellar mass range covered by our galaxies, and also in much stronger disagreement with DR3). 
In Run~4, 5 and ~6 we have included more Ti and/or Fe spectral indices for both Mg indices), without significant changes to the results.  %Finally, in Run~5 we have used the Mgb together with 4 Fe indices, and in Run 6 not fitting for [Na/Fe] to see whether we recover the metallicities calculated from the fiducial set. However, the metallicities remained consistently lower for all the runs including Mg5177. 

Secondly, for Runs~7-13, we focused on the ages and we investigated the impact of different age-sensitive features and methods. 
As already mentioned, for our fiducial fitting we have kept the age fixed to the one derived by \ppxf\ and did not use any of the Balmer spectral indices. In Run~7 and ~8 we have tested the result of including $H\beta$ while keeping the age fixed. As expected, this did not lead to any significant changes in the inferred stellar population parameters (since hydrogen spectral indices mostly impact age). 
In Run~9 we have used the same set of indices as in the previous runs, but we have left the age free to vary from 0 to 14 Gyr. This led to significant changes in the results but not on the IMF slope. 

Finally, in the remaining runs, we have tested different combinations of the above-described changes to the fiducial set of indices. Remarkably, the inference on the IMF slope is very robust and does not change by changing the considered indices.   
%In the following 2 runs (Run~10, 11) we simply repeat the above test but changing the the Balmer line used to constrain the age. However, this did not lead to any significant changes from their respective runs. 
%Finally, in Runs 12-19 we have tested the combinations of the above changes to the fiducial set of indices.  In none of these runs the results were significantly different from the fiducial fit. 
Considering also that the number of galaxies with uncontaminated indices is the greatest for the six chosen indices (see last column of Table 2), this is indeed the best choice\footnote{Removing NaD and not fitting for [Na/Fe] increases the number of galaxies by 1. However, it decreases the number of parameters constrained during the fit and might influence the inference on IMF (see \citealt{Spiniello+14}). }

\begin{figure*}
 \centering 
\includegraphics[width=1\textwidth]{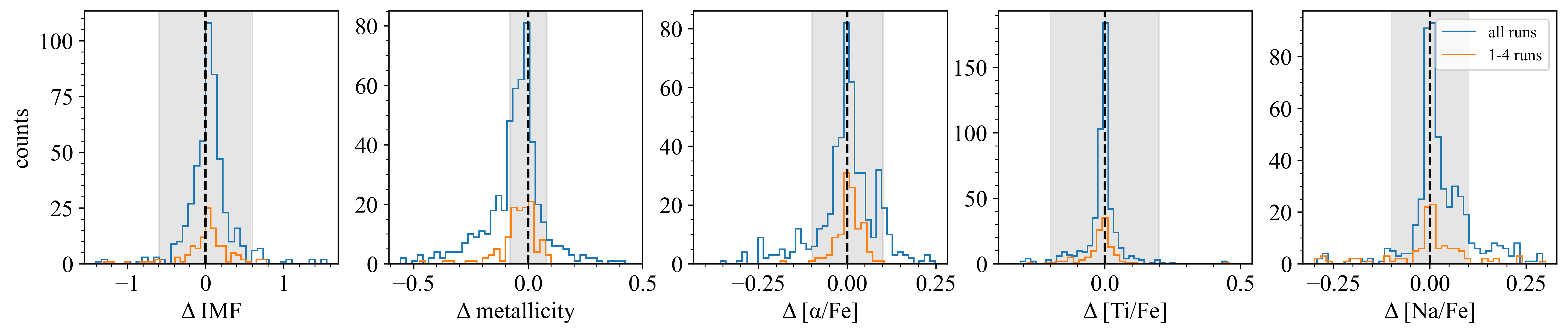}
  \caption[width=0.5\textwidth]{Histograms of the different results of all 19 FIF runs (blue) and the runs from 1 to 4 only (orange) and the fiducial fit. Each panel shows results for a different stellar population parameter. The majority of the results, as well as the overall distribution, is consistent with the fiducial fit results (within the errors, grey area), confirming the robustness of our fitting.}
  \label{fig:histodelta}
\end{figure*}

In Figure~\ref{fig:histodelta} we show, for each galaxy in \INSPIRE, the distribution of the difference between the value of each parameter inferred from the fiducial run and these inferred from the 19 different runs listed in Table~\ref{tab:runs}. Each panel shows a stellar population parameter. The results of a great majority of the runs are consistent within the errors (grey vertical regions) with the fitting with the 6 chosen indices, which demonstrates that the results are robust and overall independent from the choice of the set of indicators. 

%leads us to believe that the results would not be much different if greater number of indices was being used, thus proving the constraining power of the chosen indices. 


%Then, we investigated the differences in the obtained stellar population parameters, the result of which is shown in Figure~\ref{fig:fig3}. %\hyperref[fig:fig2]{2}. 
%The five panels show histograms of differences in the parameters between the fitting with the 6 chosen spectral indices and fitting with every other investigated combination of indices for every galaxy. 

%\begin{figure*}

% \centering 
%  \includegraphics[width=1\textwidth]{index_definition_and_response_plot.png}
%  \caption[width=0.5\textwidth]{The top panels show the Mg2 (left), Fe5270 (middle) and the NaD (right) spectral features, normalized using the index pseudo-continua (blue shaded regions). In the FIF approach, every pixel within the central bandpass (grey area) is fitted to obtain the stellar population parameters. The black line correspond to a model of solar metallicity [M/H] = 0, [Mg/Fe] = 0, [Ti/Fe] = -0.3, [Na/Fe] = -0.3 and IMF slope of $\Gamma$\textsubscript{B}. The bottom panels show the relative change in the spectrum after varying different stellar population parameters: the IMF slope was varied by $\Gamma$\textsubscript{B} = 1, the metallicity by 0.4 and abundance ratios by 0.6.}
%  \label{fig:fig1}

%\end{figure*}

\section{Comparison with INSPIRE DR3}
\label{app:comparisonDR3}

\begin{figure*}
 \centering 
\includegraphics[width=\linewidth]{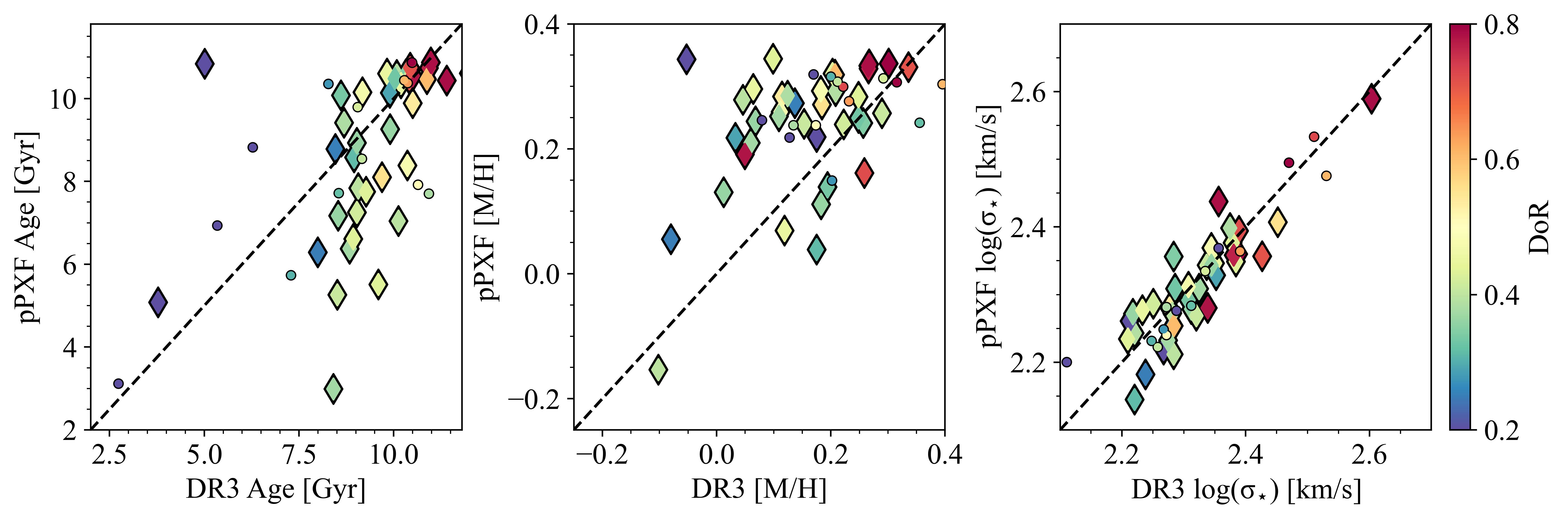}
  \caption{Comparison between the DR3 results and the ones obtained in this paper with \ppxf, colour-coded by the DoR. As in previous figures, galaxies in the \textit{Golden Sample} are shown as diamonds, although we note that these results are obtained from full-spectrum fitting, which should overcome the problem of having contamination within one or more index bandpass.}
  \label{fig:dr3comp}
\end{figure*}

%As one of the tests To confirm the robustness of our results, we  compared 
In this Appendix, we compare the ages, metallicities, and velocity dispersion values, that we have obtained by fitting the SSP models with pPXF in the first step of the analysis presented in this paper, with the results published in \citetalias{Spiniello+23}. 
We note however that a perfect agreement is not expected between these two measurements. 
We stress that, although the values are computed with the same code, the assumptions in the two cases were different. 
In fact, in  \citetalias{Spiniello+23} we kept the IMF fix to $\Gamma_b=1.3$ and performed the fit using models with a $[\alpha$/Fe] value corresponding to the [Mg/Fe] ratio inferred from line indices. Here we let the IMF vary during the fit but assume a solar $\alpha$ abundance for the models. 
%Another difference between the two measurements is that for DR3, we estimated the [Mg/Fe] via line-index analysis and then performed the \ppxf\ fit with SSP models having [$\alpha$/Fe] abundance equal to the [Mg/Fe]. Here, instead, we perform the fit with models with [$\alpha$/Fe] = 0.

Moreover, the ages we presented in DR3, are averaged values obtained from the unregularised and the maximum regularised fit. Here we do not perform any regularisation. 
Finally, another big difference in fitting the ages is that while in DR3 we worked in logarithmic space and then only converted to linear ages when plotting, in the version of \ppxf\ used for this paper, the ages are converted into linear values in Gyr before fitting. 

Indeed, as seen in Figure~\ref{fig:dr3comp}, the scatter is large for ages and metallicities\chiara{, reflecting the degeneracy between the IMF and the other stellar population parameters. This } demonstrates that one needs to take it into account, even when only limiting the fit to the optical wavelength range. 
Overall the results from DR3 favour older ages and larger metallicity values. 
We remind the reader that age and metallicity are degenerate \citep{Worthey+94}, and both quantities also correlate with IMF \citep{LaBarbera+13_SPIDERVIII_IMF, Spiniello+14, Sarzi+18,Barbosa21}. 
A very good agreement is found instead for the stellar velocity dispersion values, at all DoR. 

Nevertheless, as already highlighted in previous \INSPIRE\ papers, the star formation history inferred from spectral fitting depends much more on the parameters and assumptions for non-relics than for relics. This is true also in this case: we find a much better agreement for objects with DoR$\ge 0.6$, while the scatter increases both for age and for metallicity below this value.

%. with apart from some scatter, the points are close to one-to-one line and the two sets of results are consistent within the errors.

\section{Spectral indices definition}
\label{app:indices}
In Table~\ref{tab:indices} we provide the index definition, along with the blue and red continuum bandpasses for all the indices used in this paper. We also give the reference to the paper where each index was defined and used for the first time. 

\begin{table*}
\centering
\caption{Indices bandpasses, units in which they are measured, and papers from which they have been taken.}
\label{tab:indices}
\begin{tabular}{cccccc}
\hline\\[-1em]
\multicolumn{1}{c}{Line-index} &
\multicolumn{1}{c}{Blue bandpass (\AA)} &
\multicolumn{1}{c}{Index bandpass (\AA)} &
\multicolumn{1}{c}{Red bandpass (\AA)} & 

\multicolumn{1}{c}{Units}& 
\multicolumn{1}{c}{Reference paper}\\
\\[-1em]\hline\\[-1em]
H$\beta$  & [4815.000$-$4845.000] & [4851.320$-$4871.320] & [4880.000$-$4930.000] & \AA & \citet{Vazdekis10} \\
H$\delta$A & [4041.600$-$4079.750] & [4083.500$-$4122.250] & [4128.500$-$4161.000] & \AA & \citet{Vazdekis10}  \\
Mg2        & [4895.125$-$4957.625] & [5154.125$-$5196.625] & [5301.125$-$5366.125] & mag & \citet{Trager98}\\
Mgb5177    & [5142.625$-$5161.375] & [5160.125$-$5192.625] & [5191.375$-$5206.375] & \AA & \citet{Trager98} \\
Fe5015     & [4946.500$-$4977.750] & [4977.750$-$5054.000] & [5054.000$-$5065.250] & \AA & \citet{Trager98}\\
Fe5270     & [5233.150$-$5248.150] & [5245.650$-$5285.650] & [5285.650$-$5318.150] & \AA & \citet{Trager98} \\
Fe5335     & [5304.625$-$5315.875] & [5312.125$-$5352.125] & [5353.375$-$5363.375] & \AA & \citet{Trager98}\\
Fe5709     & [5672.875$-$5696.625] & [5696.625$-$5720.375] & [5722.875$-$5736.625] & \AA & \citet{Trager98}\\
NaD        & [5860.625$-$5875.625] & [5876.875$-$5909.375] & [5922.125$-$5948.125] & \AA & \citet{Trager98}\\ 
TiO1       & [5723.000$-$5750.000] & [5945.000$-$5994.125] & [6038.625$-$6103.625] & mag & \citet{Spiniello+14}\\
TiO2       & [6066.600$-$6141.600] & [6189.625$-$6265.000] & [6422.000$-$6455.000] & mag & \citet{Spiniello+14}\\
TiO3       & [7017.000$-$7064.000] & [7123.750$-$7162.500] & [7234.000$-$7269.000] & mag & \citet{Spiniello+14}\\
\\[-1em]\hline\\[-1em]
\end{tabular}
\end{table*}

%%%%%%%%%%%%%%%%%%%%%%%%%%%%%%%%%%%%%%%%%%%%%%%%%%

% Don't change these lines
%\bsp	% typesetting comment
\label{lastpage}
\end{document}